\documentclass[12pt]{article}

\newcommand{\mathsym}[1]{{}}
\def\id{\protect{{1 \kern-.28em {\rm l}}}}

\def\be{\begin{eqnarray}}
\def\ee{\end{eqnarray}}

\makeatletter
\renewcommand\section{\@startsection {section}{1}{\z@}%
                                   {-3.5ex \@plus -1ex \@minus -.2ex}%
                                   {2.3ex \@plus.2ex}%
                                   {\normalfont\large\bfseries}}
\renewcommand\subsection{\@startsection{subsection}{2}{\z@}%
                                   {-3.25ex\@plus -1ex \@minus -.2ex}%
                                   {1.5ex \@plus .2ex}%
                                   {\normalfont\normalsize\bfseries}}

\makeatother

\def \foot {\footnote}
\def \bi{\bibitem}
\def \tr {{\rm tr}}
\def \ha {{1 \over 2}}
\def \td {\tilde}
\def \ci{\cite}

\def \const {{\rm const}}

\def \t {\tau}
\def\S{{\mathcal S} }

 \def \J {{\mathcal  J}}

\def \d {\del}

\def\a{\alpha}

\def\C{{\bf C}}

\def \del{\partial}
\def \a {\alpha}

\def\g{\gamma}
\def\s{\sigma}

\def\ov{\over}

\def\J{{\mathcal J}}

\def\l{\lambda}

\def \k {\kappa}
\def\foot{\footnote}

\def \ci {\cite}

\def \foot {\footnote}
\def \bi{\bibitem}
\def \ha {{1 \over 2}}

\def \fo { {1\ov 4}}
\def \ep {\epsilon}

\def \d {\partial}

\def \l  {\lambda}

\def \const {{\rm const}}
\def \V {v}

\def \S {{\rm S}}

\def \td {\tilde}

\def \D {\Delta}

\def \m {\mu}

\newcommand{\bra}[1]{\mbox{$\langle #1 |$}}
\newcommand{\ket}[1]{\mbox{$| #1 \rangle$}}

\def \bi{\bibitem}
\def \la {\label}

\def \l {\lambda}
\def\foot{\footnote}

\def \sql {{\sqrt \l}}
\def \adss {$AdS_5 \times S^5~$ }
\newcommand{\rf}[1]{(\ref{#1})}
\def \ov {\over}

\def\cc{\circ}

\def \ha{{1\ov 2}}

\def\r{{\rm r}}

\def \no {\nonumber}

\def \J {\mathcal{J}}
\def \del {\partial}

\def \S {{\cal S}}
\def \J {{\cal J}}

 \def \bb {\bar \beta}

\def \bi{\bibitem}
\def \la {\label}

\def \l {\lambda}
\def\foot{\footnote}

\def \sql {{\sqrt \l}}

\def \adss {$AdS_5 \times S^5$\ }

\def \D {\Delta}

 \def \t {\tau}
 
 \def \r {\rho}

\def \ov {\over}

\def \varpi {{\rm w}}
\def \OO {{\cal O}}

\def \pa{\partial}

\def \ep {\epsilon}

\def \KK {{\rm K}}

\def \te {\theta}

\def \cc {{\rm f}}

\def \S  {{\rm S}}

\def \pa {\partial}
\def \C {{\cal C}}
\def \bea {\be}
\def \eea {\ee}

 \def \t {\tau}

\def \KK {{\rm K}}

\def \del {\partial} 
\def \d {\partial}
\def \s {\sigma}

 \def \J {{\cal J}}
 \def \S {{\cal S}}

\def \d {\del}
\def \bd {\bar \del}

 \def \tes {\textstyle}
\def \os  {\OO({\textstyle{ 1\ov \sql}})}
\def \oss  {\OO({\textstyle{ 1\ov (\sql)^2}})}
\def \osss  {\OO({\textstyle{ 1\ov (\sql)^3}})}

\def \bd {\bar \del} \def \sql {\sqrt{\lambda}} 
\def \vp {\varphi}

\def \edo \end{document}

\def \S {{\cal S}}

\def \ha {{{\textstyle{1 \ov2}}}}
\def \fo {{\textstyle{1 \ov4}}}
\def \C  {{\cal C}}
\def \sql {{\sqrt{\l}}}
\def \S {{\cal S}} 
\def \D  {\Delta }

\def \ket  {\rangle}
\def \bra  {\langle}
\def \sql  {\sqrt{\l}}
\def \V  { {\rm V}} 
\def \CC {{\rm C}}
\def \xe  {x_{0e}}
\def \bea {\be} \def \eea {\ee} \def \eqref  {\rf}

\def \rY {{\rm Y}}
\def \rX {{\rm X}}
\def \ha {{1 \ov 2}}

\def \tae {\tau_e} 

 \def \bc { {\hat c}}

\begin{document}


\overfullrule=0pt
\parskip=2pt
\parindent=12pt
\headheight=0in \headsep=0in \topmargin=0in \oddsidemargin=0in

\vspace{ -3cm}
\thispagestyle{empty}
\vspace{-1cm}

\rightline{Imperial-TP-AT-2010-3}


\begin{center}
\vspace{1cm}
{\Large\bf  

On semiclassical 
computation  of 3-point functions \\
of closed  string vertex operators 
in $AdS_5\times S^5$ 

\vspace{1.2cm}

   }

\vspace{.2cm}
 {
R. Roiban$^{a,}$\footnote{radu@phys.psu.edu} and A.A. Tseytlin$^{b,}
$\footnote{Also at Lebedev  Institute, Moscow. tseytlin@imperial.ac.uk }}\\

\vskip 0.6cm

{\em 
$^{a}$Department of Physics, The Pennsylvania  State University,\\
University Park, PA 16802 , USA\\
\vskip 0.08cm
\vskip 0.08cm $^{b}$Blackett Laboratory, Imperial College,
London SW7 2AZ, U.K.
 }

\vspace{.2cm}

\end{center}

\begin{abstract}
 We consider the  leading large string tension  correction 
 to correlation functions of three vertex operators of  particular 
  massive string states in 
 \adss  string theory. We assume  that  two  of these states are ``heavy'' 
 carrying  large spins (of order  string tension) 
 and thus  can be treated semiclassically  while 
 the  third state  is ``light''   having fixed quantum numbers. 
 We  study  several examples. In the 
  case  when the    ``heavy''  states are described 
 by a folded  string  with large spin
 in $AdS_5$   the 3-point function scales as a semiclassical spin 
 parameter  of 
 the ``heavy'' state
 in power of the 
 string level of
 the ``light'' massive string state. 
 We observe   similar behavior in the case  of 
   ``heavy" states which admit a small  angular momentum  limit  
 which may  thus  represent  
 correlators of three quantum massive  string states.

\end{abstract}

\newpage
\setcounter{equation}{0} 
\setcounter{footnote}{0}
\setcounter{section}{0}

\renewcommand{\theequation}{1.\arabic{equation}}
 \setcounter{equation}{0}

\setcounter{equation}{0} \setcounter{footnote}{0}
\setcounter{section}{0}

\def \edd {\end{document}}

\def \cc {{c }} 
\def \OO {{\cal O}}
\def \te {\textstyle}
\def \fl {\sqrt[4]{\l}}

\def \ha {{{\textstyle{1 \ov2}}}}
\def \fo {{\textstyle{1 \ov4}}}
\def \rx {{\rm x}}
\def \hg {{\hat \g}}

\def \x {\xi}

\def \C  {{\rm C}}
\def \hC  {{\rm \hat  C}}
\def \dd  {{\rm d}}
\def \bb {{\rm b}}
\def \dDelta {2}
\def \sql {{\sqrt{\l}}}

 \def \an {{\rm an}} \def \nan {{\rm nan}}
 \def \D {\Delta}
  \def \tauE {\te} \def \zx { \xi} \def \rx {{\rm x}}
  \def \ed {{\end{document} }}
  

\section{Introduction }
According to the AdS/CFT   duality 
\ci{mal}
  between the  ${\cal N}=4$ SYM theory and the  superstring theory in \adss 
  the planar  correlators of  single-trace  conformal primary operators 
   in gauge theory  should be related  to the 
correlation functions of  the corresponding closed-string vertex
operators on a worldsheet with the 2-sphere topology. The  integrated 
vertex operators may be parametrized 
by a point $\rx^m$ on the $AdS_5$  boundary $V(\rx)=  \int d^2 \xi \  
\V (x(\xi)-\rx; ...)$. They depend on quantum  numbers $Q_i= (S, J, ...)$ (such 
as spins and orbital momenta) and the 4d dimension (or AdS energy) $\D$ 
of the string  states they represent.
The  dimension $\D$  is related to the quantum numbers 
$Q_i$  and the string tension $T= {\sql \ov 2 \pi}$  by  the 
marginality condition on the vertex operator~$\V $. 

As we will review below in section 2, the 
vertex operators have generically an exponential 
dependence on the dimension $\D$ and the charges $Q_i$ of the corresponding 
string states. Thus, when these quantum numbers are as large as the  string 
tension,  the vertex operators effectively scale exponentially
 with the string tension.
It is then natural to expect that the leading  large $\sql$ contribution to 
correlation functions of such operators is  determined  by a semiclassical string 
trajectory with sources provided by the vertex operators. 
This observation may lead to a prediction for the  strong-coupling behaviour  
of  the corresponding gauge theory correlators for the dual  non-BPS operators. 

Such  semiclassical approach was developed successfully  for the calculation 
of 
two-point functions 
in 
  \ci{pol,gkp2,tsev,b1,but}
  and also for the calculation of correlators involving Wilson loops  
   \ci{bcfm,zar,yo,tsu,Sak}.  
A generalization to certain
three-point functions was discussed in \ci{ja,but}  and more 
 recently addressed in \ci{zarr,cost}.

More generally, one may consider  a  correlation function of some number 
of ``heavy'' (or ``semiclassical'') vertex operators $V_H$  with 
$\D \sim Q_i \sim \sql \gg 1$  and some number of  ``light'' (or ``quantum'') 
operators $V_L$  with $  Q_i \sim  1$ and $ \D \sim \fl$ 
(or $\D \sim 1$ for  
 ``massless'' or BPS states).
In this  case one may again expect that, in a large $\sql$ expansion, the 
leading order contribution to
\be   \KK_{H_1... H_n L_1...L_m }= \bra V_{H_1}(\rx_1)  ... 
 V_{H_n} (\rx_n) \   V_{L_1} (\rx_{n+1} ) ... V_{L_m}(\rx_{n+m} ) \ket   \la{aaa} \ee  
will be  given by the  semiclassical string trajectory determined 
by the ``heavy''  operator insertions.
To  compute $ \KK_{ H_1... H_n L_1...L_m }$  one should first construct the classical 
string solution that determines the leading large $\sql$ contribution to 
$
\KK_{H_1...H_n}= \bra V_{H_1}(\rx_1)  ... V_{H_n} (\rx_n)\ket  $
and then compute  \rf{aaa}   by simply evaluating the product of 
``light" vertex operators $V_ {L_1}(\rx_{n+1} ) ... V_ {L_m}(\rx_{n+m} )$ 
on this solution.
 
One  may understand  this procedure as a limit of the general semiclassical 
computation  for the correlator of $n+m$ ``heavy'' operators, all of which have 
large quantum numbers.  
In this case the classical  trajectory should be determined  by a solution of 
the string equations with source  terms  provided by {\it all} the $n+m$ operators. 
Finding such surface appears to be 
 hard in general, but if we formally assume  
that the charges of $m$ of the $n+m$ sources are much smaller than the 
other $n$,  then the semiclassical trajectory will be dominated 
by the contribution 
of the $n$ large charges
 (the effect of the $m$ small-charge sources may then be included
perturbatively).
 Thus the  leading contribution to the correlator  will then be
computed as suggested above for $ \KK_{H_1..H_n L_1... L_m}$. 
We will return to the discussion of the 
 validity of this approach 
and   the 
computation of quantum ($1 \ov \sql$) 
 corrections to the leading approximation 
 in the concluding section 5.

Three-point correlation functions are the first nontrivial examples where 
these considerations  become  relevant.  
%
%
 While finding a semiclassical trajectory  controlling the leading 
 contribution  to 
 $\bra V_{H_1}(\rx_1) V_{H_2}(\rx_2) V_{H_3} (\rx_3)\ket $
 is so far an unsolved problem \ci{ja}, 
 the discussion above suggests that one can use the semiclassical trajectory 
 for the correlation function of {\it two}   ``heavy'' operators  
  $\bra V_{H_1}(\rx_1) V_{H_2}(\rx_2)\ket $,  which  is 
 straightforward to find  \ci{tsev,but},
 to compute the leading contribution to a  correlator containing two
 ``heavy'' and one ``light''   state
  $\bra V_{H_1}(\rx_1) V_{H_2}(\rx_2) V_{L}(\rx_3)\ket $.
 Examples of such computations,   with $V_H$ 
 corresponding to a semiclassical  string state with large spin in $S^5$  
 and $V_L$ representing a 
 BPS state corresponding to a massless (supergravity) 
 scalar  or dilaton  mode, were recently presented in \ci{zarr,cost}.

\

The aim of the present paper is to  consider more general cases 
when $V_L$  may  represent a massive string mode.\foot{Correlation
 functions 
 of (non-near-BMN) massive string states 
were not discussed in the past, apart from not directly related
   study of decay of semiclassical spinning string in   \ci{ple, cjjk}.}
We shall study few  
 explicit examples, attempting to  clarify  the general 
 structure  of  such  3-point functions.
We shall also consider several choices for the ``heavy'' operator $V_H$. 
One will be the physically interesting 
case  when $V_H$ represents a folded  string with 
large spin $S$  in $AdS_5$ 
dual to twist $J$  operator. 
We will also  try  to shed light on the correlation function  
of three massive string states  from the first excited string level 
by choosing $V_H$ to represent a ``small'' semiclassical string 
that  admits a smooth fixed  spin limit as proposed  in \ci{rt9}.

The  two-point and three-point correlation functions  are special in that their 
dependence on the position of the operators is controlled by the
 target space conformal invariance
\foot{Here we assume for simplicity 
 that  the primary operators are scalar.
 In the case of  primary  with spin operators there are extra 
  kinematic factors (see, e.g.,  \ci{osbp}).    
  For example, 3-point function  of two scalar operators 
  and one spin $s$ operator $V_{m_1...m_s}  (\rx_3)$ (which is a symmetrised
   traceless  tensor) 
   we get  an extra   factor $d_{m_1} ... d_{m_s}$, where $d_m= 
   { (\rx_3-\rx_1)_m \ov  (\rx_3-\rx_1)^2} -{ (\rx_3-\rx_2)_m \ov  (\rx_3-\rx_2)^2}$.
 In the case of ``heavy'' operators with spins such factors may be ignored as we will
 consider ratios  of 3-point functions to their  2-point functions. 
 In the case when the ``light'' operator  corresponds to  spin $s$ 
 operator like $\tr ( \bar Z D_+^s Z)$   we shall implicitly assume  
 that the corresponding extra factor  
 $ (d_+)^s =  \big[ { (\rx_3-\rx_1)_+ \ov  (\rx_3-\rx_1)^2} -{ (\rx_3-\rx_2)_+ \ov 
 (\rx_3-\rx_2)^2}\big]^s $  is included.
 }
\be 
&&\bra  V_{1} (\rx_1) V_{2} (\rx_2)\ket = 
 { \CC_{12}\  \delta_{\D_1,\D_2}  \ov   |\rx_1-\rx_2|^{2 \D_1} }   \ , 
\la{re}
 \\
&&   
\bra V_{1}  (\rx_1) V_{2}  (\rx_2) V_3 (\rx_3) \ket = 
{ \C_{123}\ov 
  |\rx_1-\rx_2|^{\D_1 + \D_2 - \D_3 } |\rx_1-\rx_3|^{ \D_1 + \D_3 - \D_2 } 
 |\rx_2-\rx_3|^{\D_2 + \D_3 - \D_1 } }     \ . \la{ko}
\ee
Here in \rf{re} $ V_{1}= V_2^*$. 
The two-point function coefficient $\C_{12}$ may be set to unity by a choice 
of normalization of vertex operators.
The three-point function coefficient $\C_{123}$ may be extracted by setting 
$\rx_1,\rx_2,\rx_3$ to specific values. As we shall see below, in the case of 
$\bra  V_{H} (\rx_1) V_{H} (\rx_2) V_L (\rx_3 ) \ket$ a natural choice will be
$\rx_1=-\rx_2, \ |\rx_1|=|\rx_2|=1\ $ (i.e. $ |\rx_1-\rx_2|=2  $)
 and  $\rx_3=0$. 
%
%

To isolate  the issue of normalization of operators  one may consider 
 ratios of  particular 3-point correlators 
 with different operators or different  values
  of quantum numbers of the same operator. Combining 
  such correlators  one may hope to extract  information about 
  normalization-independent data, like 
 factors  involving   quantum numbers of the 
 different 
 types  of the vertex operators at the same time.  For  example,  in the combined 
 ratio 
 \be
 \frac{\bra V_{H}  (\rx_1) V_{H}  (\rx_2) V_{L} (\rx_3) \ket }
          {\bra V_{H}  (\rx_1) V_{H}  (\rx_2) V_{L'} (\rx_3) \ket }
 \times
 \frac{\bra V_{H'}  (\rx_1) V_{H'}  (\rx_2) V_{L'} (\rx_3) \ket }
          {\bra V_{H'}  (\rx_1) V_{H'}  (\rx_2) V_{L} (\rx_3) \ket }
 \ee
 the normalization factors of both ``heavy" and ``light" states cancel
  out. Here 
 $H$ and $H'$  and well as $L$ and $L'$  may differ by,  e.g., 
  choice of charges.
 This ratio is determined completely by terms in the three-point function 
 ${\bra V_{H}  (\rx_1) V_{H}  (\rx_2) V_{L} (\rx_3) \ket }$ 
 which depend  in a nontrivial way on the charges of both the ``heavy" and 
 the  ``light" states. 
 
 
 \
 
The structure of the rest of the paper is as follows. 
In  section 2  we review 
the structure of the bosonic part of some 
closed string  vertex operators of the \adss superstring.
We consider  several examples which will be used  in later
sections:   the ``massless'' operators 
representing  dilaton  and 
 the superconformal primary state of charge $J$,
 the  massive state 
with spin $S$ on the leading Regge trajectory and a special singlet  string 
state existing  on massive  string levels. 
%

In section~3 we review the semiclassical calculation of two-point correlation 
functions of large charge operators. We discuss in detail the string states dual 
to large spin twist-2 operators and   to large twist $J$  operators.

We then proceed in section~4 to discuss the three-point functions of one ``light"
and two ``heavy'' operators discussed in sections~2 and~3. We also 
 use the same 
approach to construct the three-point functions in the case when  the ``heavy''
operators are described by a classical trajectory admitting a small spin  limit.
In all cases we will identify the normalization independent features of the 
three-point function coefficient. 

Some concluding remarks including  comments on the validity of our approach 
and on the calculation of quantum corrections to the three-point functions
are made  in section~5.


\renewcommand{\theequation}{2.\arabic{equation}}
 \setcounter{equation}{0}
 \section{Examples  of string vertex operators }
 
Let us start  with a review of the structure of relevant vertex operators 
following \ci{tsev,rt9}. Their form is perhaps most transparent in 
the $6+6$~embedding coordinates.\foot{We shall follow the notation of \ci{but}.
The relation to the notation for the coordinates of $AdS_5$ and $S^5$ 
in \ci{tsev} is: $Y_M \to N_M$, \ $X_k \to n_k$.} In these coordinates the action
of the  \adss superstring sigma model  has the following structure:
%
%
\bea 
\la{scc} && I= { \sql \ov 4 \pi}  \int d^2 \xi
 \ \Big(  \del Y_M \bd Y^M +  \del X_k \bd X_k 
+ {\rm fermions\ } \Big)  \ , \\
&&  Y_M  Y^M =- Y_0^2 - Y_5^2  + Y_1^2 + Y_2^2 +  Y_3^2 +Y_4^2 =-1  , 
\ \ \ \ \ \ 
 X_k X_k = X_1^2 +...+ X_6^2=1   ~~~  \la{nen} 
\eea
For a worldsheet with Minkowski signature the 2d derivatives are
$\del = \del_+, \ \bd = \del_-$.
%
%
In general, the vertex operators are  constructed in  terms of $Y_M$,  $X_k$ 
and fermions and correspond  to the highest weight states  of 
$SO(2,4) \times SO(6)$ representations. 
They are (exactly) marginal operators of dimension 2, i.e.  are particular linear 
combinations  of products of $Y_M, X_k$ and their derivatives that are  
eigenvectors of the 2d anomalous dimension operator.\footnote{Even though  the sigma model of the Green-Schwarz type 
is not a factorized CFT, the (anti)holomorphy of the two components of 
the stress tensor guarantees that the left and right dimensions are 
well-defined quantities.}
%
%
Fermions render the \adss  sigma model UV finite; since we will be 
interested in the leading order of the semiclassical expansion we
may  nevertheless ignore them both in the action and  the vertex operators. 
Being interested in the leading large string tension 
 approximation we may also 
ignore all $\a' \sim  { 1 \ov \sql}$ corrections to 
the bosonic part of the vertex  operators (see also  section 5). 

Let us  recall  the basic relation between the embedding coordinates and the 
global and Poincar\'e coordinates in $AdS_5$  that we will use  below:
\be
&&Y_5 + iY_0 =\cosh \rho\ e^{it}\,, \quad 
Y_1 +i Y_2=\sinh \rho\ \cos\theta\ e^{i \phi_1}\,, \quad
Y_3 +i Y_4=\sinh \rho\ \sin\theta\ e^{i \phi_2}\,,
\nonumber \\
&& Y_m=\frac{x_m}{z}\,, \qquad
Y_4=\frac{1}{2 z}(-1+z^2 +x^m x_m) \,, \qquad 
Y_5 = \frac{1}{2 z} (1+z^2 +x^m x_m)\,   , 
\label{yy}
\ee
where  $  x^m x_m = - x_0^2 + x_i x_i$ \  ($m=0,1,2,   3; \ i=1,2,3$).
If a highest-weight state of an $SO(2,4)$ representation  is labelled by the three  
Cartan generators $(E,S_1,S_2)$ corresponding to rotations in the 
planes $(5,0)$, $(1,2)$ and $(3,4)$,  a  wave function or a vertex 
operator representing a state  with  AdS energy  $E$  should   contain 
a factor
$ (Y_5 + iY_0)^{-E} = (\cosh \rho)^{-E} \  e^{-iEt} $.  This is just the 
AdS analog of the flat-space energy dependent plane wave factor $e^{-iEt} $.
If, equivalently,  the representation is  labelled by the $SO(1,1)$ generator in 
the $(5,4)$ plane, then the corresponding factor is
$  (Y_5 + Y_4)^{-\D} $,
where  $\D$ is the eigenvalue of the 
dilatation generator  (acting as $z\to  k z,\ x_m \to  k x_m$).

For the construction of the classical string solution describing
 the semiclassical approximation of the  two-point 
function of  ``heavy" vertex operators it is useful, as described in \ci{tsev,but},
to  consider the  euclidean continuation  
\be \la{p}
t_e=i t  \ , \ \ \ \ \ \    Y_{0e}= i Y_0 \ , \ \ \ \ \ \ x_{0e} = i x_0 \ , 
\ee
so that $ Y^M Y_M= -Y_5^2 +Y_{0e}^2   + Y_i Y_i  +Y_4^2 =-1$.
The  $SO(2,4)$ symmetry  is then replaced by $SO(1,5)$, which  contains  
the discrete  transformation 
$
Y_{0e} \leftrightarrow Y_4 , \  \ E \leftrightarrow \D 
$ 
that relates the factors  $(Y_5 + iY_0)^{-E} $ and $  (Y_5 + Y_4)^{-\D} $.
Up to a normalization factor, we shall
sometimes  denote this factor  by $K$ in the 
following:
\be 
 &&  K(x,z) =k_\Delta\ ( \rY_+)^{-\D} 
              = k_\Delta\ \left(z+ z^{-1} x^m x_m\right)^{-\Delta}~~, \ \ \ \ \ \ \ \ 
	\rY_+ \equiv Y_5 + Y_{4} \ , \\
	&&
k_\Delta= { \Gamma (\D) \ov \pi^{d/2} \Gamma (\D- {d\ov 2})}\Big|_{_{d=4}} 
= {1 \ov \pi^2} (\D-1) (\D-2) 	 \ .  \no      
 \la{gh} 
 \ee 
 As is well known, 
 $K(x-x',z) = k_\D \left[z+ z^{-1} (x-x')^2 \right]^{-\Delta}$
 is a solution of the scalar Laplace equation in $AdS_5$ with mass 
 $m^2 = \D (\D-4)$; the normalization constant 
 is chosen such that $K(x-x',z\to 0) = \delta^{(4)} (x-x')$. 
 
In general, an unintegrated vertex operator will have the structure  
\be \V \sim (\rY_+)^{-\D} \ \Big[ (\del^s Y)^r ... (\bd^m X)^n + ... \Big]
\equiv (\rY_+)^{-\D}\ U (Y,X,...) \la{stuu} \ . \ee
To construct an  integrated  vertex operator parametrized  by the 
four coordinates of a point on the boundary of the  euclidean Poincar\'e 
patch of $AdS_5$  space, we should  shift $x_m =(\xe,x_i)$ by a constant 
vector $\rx_m $ 
(translations in $x_m$ are part of global 
conformal symmetry)
\be 
 V(\rx)= \int d^2 \xi \  \V \big( x(\xi) -\rx;\ ... \big) =  \int d^2 \xi \  
 K(x(\xi)-\rx,z(\xi))
  \ U[x (\xi)-\rx,z(\xi), X(\xi)]  
  \ .   \la{jk}
\ee
Let us now discuss some examples of  such vertex operators which 
we shall  use as 
``heavy'' or ``light'' factors in the 3-point correlation functions
below.

 \subsection{Dilaton operator}
The 10-d dilaton field is decoupled from the metric perturbation 
in the Einstein frame \ci{pvn}, i.e.  it satisfies the free massless 10-d 
Laplace equation  in \adss. 
Keeping non-zero value of $S^5$ momentum (corresponding to a 
higher KK harmonic of the 10-d dilaton), the corresponding massless
string vertex operator representing a highest weight state of 
$SO(2,4) \times SO(6)$ is simply proportional to the worldsheet   
Lagrangian 
\be \la{vew}
 \V^{(\rm dil)}_J   & =&   (\rY_+)^{- \D}\  (\rX_x)^J  \ \big( \del Y_M \bar \del Y^M
   + \del X_k  \bd X_k  + 
   {\rm fermions}\big)
    \ ,  \\
    \rX_x &\equiv& X_1 + i X_2 = \cos  \vartheta \  e^{i \vp}  \ . \no
    \ee
Here and below  in this section we shall ignore the fermionic 
terms and overall normalization factors in the vertex operators.
%
The  marginality condition is 
  $ 2= 2   - { 1 \ov 2\sql} \Big[ \D(\D-4) - J(J+4) \Big]  + \oss  $,  
so that to the leading order in the large $\sql$ expansion $\D=  4 + J  $.
Inclusion of fermions should guarantee that  all  higher-order corrections 
vanish as this should be a BPS state.  
The corresponding  dual gauge theory operator 
should be  $\sim \tr ( F^2_{mn} Z^J + ...)$ or, for $J=0$,  just the  SYM 
Lagrangian.

The form of the resulting  integrated dilaton operator \rf{jk}, \rf{vew} 
can be understood as follows. On string side, 10-d dilaton couples 
to string action as 
$\int d^2 \xi \  e^{ {1 \ov 2}  \Phi(X)} g_{IJ}(X)  \del  X^I \bar \del X^J +
...$
where $g_{IJ}(X)$  is Einstein-frame metric  and $X^I$ are 10-d coordinates. 
To get the on-shell (marginal)  vertex operator one is to  linearize in $\Phi$ 
and restrict $\Phi$ to be  a solution of the 
corresponding wave  equation. In the AdS/CFT context we should then  have 
(ignoring KK momentum dependence) 
$\Phi(x,z) = \int d^4 \rx\  K(x-\rx, z) \ \phi(\rx)$, 
where the ``4-d dilaton'' $\phi(x)= \Phi(x,z\to 0)$ is an arbitrary boundary source
function. 
 The corresponding (D3-brane) coupling on the gauge theory
side is $\int d^4 \rx \ \tr \big[   e^{ - \phi (\rx)}  F^2_{mn}(\rx) + ...\big]$.
The string-theory and  gauge-theory correlation functions are then obtained by 
taking functional derivatives over $\phi(x)$;
insertion of the gauge theory Lagrangian into a gauge-theory correlator 
corresponds to insertion of $V^{(\rm dil)} $, i.e. the 
string theory Lagrangian {\it multiplied}  by the function $K \sim  (\rY_+)^{- \D}$, 
into the  string-theory correlator. 

Note that   the constant part of the dilaton  appears in  the string  action
in the same way as the string tension factor $\sql$  and  in the  gauge theory 
action as the gauge coupling $\l$. 
Taking  the derivative $ \l { \del \ov \del \l }$ of a gauge-theory correlator 
 corresponds  to the insertion of the gauge-theory action; applying 
 $ \l { \del \ov \del \l }$ to a string-theory correlator 
 corresponds  to the insertion of the string-theory action.
 The two  are indeed related as the 
 ``zero-momentum dilaton'' corresponds  to the dilaton 
 operator ($\Delta=4$) integrated over the 4-space, 
 \be \la{zdil} 
 V^{(\rm {0-dil})} \equiv   \int d^4  \rx  \  V^{(\rm dil)} (\rx) \to 
 \int d^4  \rx  \ \int d^2 \xi \  
 \big(z + z^{-1} |x(\xi) -  \rx|^2\big)^{-4}\ \big( \del Y_M \bar \del Y^M + ... \big)
 \ee
   Doing  first  the integral over $\rx$ 
 one finds that the $K \sim \big(z + z^{-1} |x -  \rx|^2\big)^{-4} $
 factor goes away and we end up  just with the string action, i.e. 
 $ V^{(\rm {0-dil})} \sim \int d^2 \xi \  \big( \del Y_M \bar \del Y^M + ... \big)$.
 
This implies, in particular, the following 
``zero-momentum dilaton'' relation, 
\be 
\la{diz}
\bra V (\rx_1) V^*(\rx_2) V^{(\rm 0-dil)} \ket=
 \l { \del \ov \del \l }  \bra V(\rx_1) V^*(\rx_2) \ket 
= - \l { \del \Delta \ov \del \l }  \   { 1\ov |\rx_1- \rx_2|^{2 \Delta} } \ln |\rx_1
- \rx_2|^2 \ ,  
\ee
i.e. the insertion into a 2-point function 
of the dilaton operator integrated over 4-space (i.e. of the gauge theory 
action on the  gauge theory 
side  or the string theory action on the string theory side) 
is proportional to the $\l$ derivative of the dimension (see  \ci{cost}
for a closely related discussion). 

As a result, the  $\C_{123}$ 
corresponding to $\bra V (\rx_1) V^*(\rx_2) V^{(\rm dil)}(\rx_3) \ket$ 
 should be proportional to $\l { \del \ov \del \l } \Delta$. Indeed, 
taking $\Delta_1=\Delta_2=\Delta$ and $\Delta_3=4$ in  \rf{ko}
and integrating over $\rx_3$ one  gets 
$ \C_{123} |\rx_1- \rx_2|^{-2 \Delta + 4}
  \int d^4 \rx_3 \  { 1 \ov  |\rx_3 - \rx_1|^4  |\rx_3 - \rx_2|^4  }$.
  The latter 
   is proportional to 
$ \C_{123} |\rx_1- \rx_2|^{-2 \Delta} \ln  ( \ep |\rx_1 - \rx_2|) $ 
\ ($\ep$ is a  cutoff) and should be compared with \rf{diz}.

\subsection{Superconformal primary  scalar operator}

This scalar represents the  superconformal primary state and is the highest weight
state  of the $SO(6)$ representation $[0,J,0]$, \  $J\geq 2$. The 
corresponding dimension  is $\D=J$.  The dual gauge theory operator 
is the BMN operator  $\tr Z^{J}$. The dilaton operator is 
the supersymmetry descendant of this operator. 

The  corresponding  massless 
string  state originates from the trace of the  graviton  in $S^5$ directions 
that induces also the  components of the graviton in $AdS_5$ directions and 
mixes with the  RR 5-form \ci{pvn,Lee}. 
As discussed in \ci{bcfm,zarr},  the  bosonic part of   corresponding  vertex operator 
can be taken in the form (ignoring derivative terms that will 
not contribute to the computation done in section 4)
\be 
\V^{(scal)}_J=  (\rY_+)^{-\D}\  \rX_x^J \ 
  \Big[  z^{-2} (\del x^m \bar \del  x_m   -  \del z   \bd z )
     -  \del X_k    \bd  X_k \Big]    \  .  \la{kz} \ee  
The two-derivative  factor here can also be written as  
$\Big[  z^{-2} (\del x^m \bar \del  x_m   -  \del Z_k    \bd  Z_k)  \Big]$,  with   
 $ Z_k = z X_k  , \  Z_k Z_k = z^2 $, so this  is 
 just  the string Lagrangian with the 4d and 6d parts 
 taken with opposite sign.\foot{Note that a similar  
 factor would appear if one would start 
with a near-horizon limit of the D3-brane metric 
$ds^2 = H^{-1/2}(z)  dx_m dx^m + H^{1/2}(z) dZ_k dZ_k, \ \   H= { Q\ov z^4}$  
and formally consider a local  deformation of the $Q $-parameter. 
A  similar    deformation (but with different coefficients for the ``4d'' and ``6d''
parts of the metric)  corresponds to a   fixed scalar 
dual to $\tr F^4+...$ operator which is a supersymmetry descendant of the 
$\tr Z^4$  operator \ci{ca}.}

\subsection{Operators with spin on leading Regge trajectory }

In flat-space (bosonic) string theory a spin $S$ state on the 
 leading Regge trajectory 
is represented by 
$
V_S= e^{ -i E t } \big( \pa \rx_x\bar{\pa} \rx_x \big)^{S\ov 2}$, \ 
$ \rx_x = x_1+ix_2$, with the marginality  condition   being 
$ 2=S  - \ha \a'   E^2$,  i.e. $E = \sqrt{{2\ov \a'} (S-2)} $.
By analogy, in $AdS_5\times S^5$,
  candidate operators for states on the leading Regge trajectory are 
  (after the euclidean continuation and $E \to \D$ flip) 
 \be 
 &&\V_S =  (\rY_+)^{-\D}\big( \pa \rY_x \bar{\pa} \rY_x \big)^{S\ov 2}+...\ ,
 \ \ \ \ \ \ \ \ \   \rY_x= Y_1 + i Y_2 \ , \la{mml}\\ 
  &&  \V_J  =  (\rY_+)^{-\D}\big( \pa \rX_x \bar{\pa} \rX_x \big)^{J\ov 2} +...\ , 
  \ \ \ \ \ \  \ \  \ \rX_x= X_1 + i X_2 \ , 
 \la{mmll}
\ee
where the ellipsis stand for terms resulting from  the diagonalization of the 
2d anomalous  dimension  operator.
In general, ignoring  fermions,   the  operator 
$ \big( \pa \rX_x \bar{\pa} \rX_x \big)^{J\ov 2}$  in  the $SO(6)$ sigma 
model may mix  with 
\be   (\rX_x)^{2p + 2q  } (\del \rX_x)^{{J\ov 2 } - 2p}   (\bar
\del \rX_x)^{{J\ov 2 }  - 2q }
( \del X_\ell \del X_\ell)^p ( \bd X_k \bd X_k)^q \ , \la{cn}   
\ee
where  $ p,q= 0,..., {J\ov 4} ; \    l,k=1,...,6$.
The  operator 
$(\rY_+)^{- \D}\big( \pa \rY_x \bar{\pa} \rY_x \big)^{S\ov 2}$  
in the  $SO(2,4)$  sigma  model  may mix  with  
\be \la{cha}
(\rY_+)^{- E - p-q} \rY_x^{p+q} (\d \rY_+)^p (\d \rY_x)^{{S\ov 2}-p} 
(\bd \rY_+)^q (\bd \rY_x)^{{S\ov 2}-q} 
+     O(  \del Y_M \del Y^M \bd Y_K \bd Y^K)  \ , 
\ee
where 
$ p,q= 0,..., {S\ov 4}  ; \   M,K=0,1,...5$.
The   true vertex operators are   
eigenvectors of the  anomalous dimension matrix, i.e.  they are
particular linear combinations of the above structures
determined, e.g.,    by solving Laplace (or Lichnerowitz)  type equation 
for the corresponding tensor wave function, e.g., 
$ \hat \gamma  \Psi = \Big[2- S +  \ha \a'  \nabla^2   + 
\sum c_k \a'^k (R....)^n...\nabla^p   \Big] \Psi =0$.

  
  
Since all operators in eqs.~(\ref{cn}) and (\ref{cha}) have the same 
classical dimension, their mixing is not suppressed by $\a' \sim {1 \ov \sql}$.
However, considering such operators  as the ``heavy'' ones in a correlation function 
(i.e. treating them semiclassically assuming that their dimension $\D$ and 
spins are as large as  $ \sql$) makes it
unnecessary to consider 
explicitly the effects of the mixing. 
Indeed,  all what is  required  is that  the classical solution they source 
 should 
have a definite  energy or  $\D$, 
 thus effectively representing 
 an eigenvector of the 2d anomalous dimension operator \ci{tsev,but}.

\subsection{Singlet scalar operators  on higher  string levels}

There exist special massive string state vertex operators 
with finite  quantum numbers  for which the leading-order 
bosonic part is known 
explicitly 
 and thus they can be used as candidates for ``light'' vertex
operators in the  semiclassical computation of the correlation functions
discussed in the introduction.
These are   singlet operators that  do not mix  with other operators
  to leading 
nontrivial order in $1 \ov \sql $   \ci{tsev,rt9}.
   
Consider, e.g.,  an operator built out of derivatives of $S^5$ coordinates 
$X_k$.  An example of a scalar operator carrying no  spins is\foot{
The marginality condition for  this operator is 
 $0=\hg =  2- 2q   +   { 1 \ov 2\sql} \Big[   \D (\D-4) +  2q(q-1)\Big]  
  +  { 1 \ov (\sql)^2 } \big[{\te{ 2 \ov 3}} q (q-1) (q- {\te{7 \ov 2}})  +
 4q \big]  + \osss$.}
 \be   \V_q= (\rY_+)^{- \D} \Big[(\d X_k \bd X_k)^q+ ... \Big] \ . \la{kop} \ee
This  operator  corresponds to a scalar string state at level $n=q-1$ 
so that the fermionic contributions   should make the  $q=1$ state massless (BPS), 
with $\D=4$  following from the marginality  condition. 
The $q=2$ choice   corresponds to a scalar state 
on the first excited string level.\foot{Then \ci{rt9} 
$\D(\D-4) = 4 \sql - 4  + \os $, so that 
$\D= 2 \fl  + 2 + { 0 \ov \fl} +  \OO({1 \ov (\fl)^3})$.
Here the   subleading terms  should   not, however, be trusted as
fermions are expected to change
the $\D$-independent  terms in the 1-loop anomalous dimension.}
 
The number of $(\d X_k \bd  X_k)$  factors in an operator
cannot increase due to renormalization~\cite{tsev}; thus 
if  an operator does not contain  any such factors, they cannot be
induced by renormalization.  
This leads to an example of another scalar   operator   which is a 
true singlet and is known explicitly at the leading 
order  
\be \la{sii}
  {\V}_r = (\rY_+)^{- \D}  (\d X_k \d X_k \bd X_\ell \bd X_\ell)^{r/2} \ , \ \ \  \ \
  r=2,4,... \
  .  \ee
 Ignoring  fermionic contributions, its  dimension  is determined from 
 $0=   \hg= 2- 2r  +   { 1 \ov 2\sql} \Big[ \D (\D-4) +  8r \Big]  
     +  \oss $, i.e. 
$\D= 2 \sqrt{  r-1 } \fl     + 2 - { 2r-1 \ov \sqrt{ r-1}  \fl}
 +  \OO({1 \ov (\fl)^3})$.
While the contribution of fermions  may, of course,   change the 
subleading terms, cf. \ci{rt9,kaz}, they cannot alter the form of the 
fermion-independent part of the vertex operator.
As this operator represents a singlet scalar, the corresponding field 
 should satisfy a simple  Laplace-type 
equation\foot{To leading order in large $\sql$ we may ignore a constant shift in
$\D$, 
i.e. ignore position of that scalar in a supermultiplet.} 
$ ( - \nabla^2 + M^2) \Phi=0, \ \ \ M^2 = \D(\D-4) = 4 ( r-1) \sql  +...$. Thus 
adding $S^5$  KK  momentum is  straightforward  by simply including a factor 
of  $\rX_x^J$  as in \rf{vew}. 

We may also  consider the $AdS_5$ counterpart of the singlet operator \rf{sii}, namely 
\be \la{siii}
  {\V}_k = (\rY_+)^{- \D}  (\d Y_M \d  Y^M \bd Y_K \bd  Y^K)^{k/2} \ , \ \ \  \ \
  k=2,4,... \
  .  \ee
The operators  $ {\V}_r $  in \rf{sii} and $ {\V}_k $ in \rf{siii}  
 have a very  special
structure:  their  derivative factor  is constructed out   of  chiral components 
$T_{++}=T$ and $T_{--}=\bar T$ of the  stress tensor of the $S^5$ or  $AdS_5$
sigma models, respectively, i.e. ${\V}_r =(\rY_+)^{- \D} (T  \bar T)^{r/2} $. 
Thus, when evaluated on a classical string solutions\foot{In conformal gauge 
the classical  stress tensors of the bosonic $AdS_5$ and $S^5$ sigma models are separately  
conserved and  traceless so that their holomorphic components can be chosen 
to be constant; the Virasoro condition  equates their sums to zero.} 
the factor $(T  \bar T)^{r/2} $ will   simply  be a constant  in power $r/2$. 
Up to this  constant   the contribution of this singlet  operator to a three-point  
correlator  with two ``heavy'' operators will  then  be the same as that of the 
``naive''  scalar operator $(\rY_+)^{-\D} \rX_x^J$.

The  simplest example of the operator \rf{sii} is $r=2$  representing 
a massive   state  on the first excited string level, 
which should be  dual to a member of Konishi multiplet (see \ci{rt9}). 
We  may thus  use  it not only  to evaluate  3-point  correlators of a singlet
massive string mode with two ``heavy'' modes represented by large spin 
operators like \rf{mml}, but also with two ``heavy'' modes corresponding to  
a ``small''  semiclassical string  which may also  be used to represent  string 
modes on the first excited level  as discussed  in \ci{rt9}. 
As we shall discuss  below, such a calculation may
 then  determine the leading term in the   correlation function of  there 
  ``light''  massive 
string modes like   three Konishi-type  states.

\renewcommand{\theequation}{3.\arabic{equation}}
 \setcounter{equation}{0}

 \section{Review of semiclassical  approximation  for 
 2-point correlator of large spin states
  }
 
The semiclassical  calculation of the correlation function
of two ``heavy'' string states  represented by large orbital momentum  
\rf{vew},\rf{kz} or large spin \rf{mml},\rf{mmll}  vertex operators 
was described  in \ci{but} (see also \ci{tsu,ja} for related discussions).
Here we review the main points of this calculation.

The classical solution describing a point-like string  with  large orbital 
momentum in $S^5$ corresponding,  e.g.,  to BMN-type  states with vertex 
operators as in \rf{vew} or \rf{kz}  with $\D, J \sim \sql$ is
$t= \kappa \tau$  (in $AdS_5$)   and $ \vp= \k \tau$  (in $S^5$).  It 
represents a massive geodesic in $AdS_5$, running through the  center of the space
and never reaching the boundary. 
After a Euclidean continuation  the 
geodesic reaches the boundary: in Poincar\'e coordinates it is (cf. \rf{p})\foot{The  
euclidean stationary point  solution for the coordinates of $S^5$ is,  in 
general,  complex  (see also \ci{pol,zar,yo,tsev,b1})  but there is no a 
priori condition that such  solution should be real.}
\be 
&&z=\frac{1}{\cosh(\kappa \tau_e)}\,, \qquad \xe =\tanh(\kappa \tau_e)\,,
\ \ \ \qquad
x_i=0\,, \ \ \ \ \vp = -i \k \tau_e \ , 
\ \ \ \ \ \  \ \tau_e=i \tau
\label{ge}  
 \ee
The radial coordinate $z$ vanishes in the limits  $\tau_e \to \pm \infty$, 
implying indeed that the euclidean trajectory 
reaches the boundary at the two points:  $\xe=-1,\ x_i=0$ and  $\xe=1,\ x_i=0$.\foot{By a dilatation and translation, the position of the two end-points
may be chosen to be $\xe=0$ and $\xe=a$;  the corresponding solution is 
then \ci{but}: \ 
$
z=  \frac{a}{ 2\cosh(\kappa \tau_e)}\,,  \  \ \  \xe =\frac{a}{2} \big[
\tanh(\kappa \tau_e)+  1 \big] 
\,, \ \ 
x_i=0. $
\la{foott}}
These points are the locations of the two vertex operators sourcing the classical 
trajectory.

Quite generally, the two vertex operators whose two-point function we are 
computing are placed at $\tae=\pm \infty$  on the Euclidean 2d worldsheet 
cylinder. Their positions may be mapped to arbitrary positions $\zx_1$ and 
$\zx_2$ on the $\xi$ complex plane  \ci{tsev,but} by  the transformation:
%
 %
 \be 
e^{\tau_e + i \sigma} = { \zx - \zx_2 \ov  \zx - \zx_1}  \ .  \la{pu} 
\ee 
Given a classical solution with given global charges on a Lorenzian 2d cylinder, 
its 
analytically continued form mapped onto the complex plane 
should then be the  stationary trajectory  of the path integral 
representing the  two-point correlation function of the vertex operators
with the given global charges. The ``delta-function'' sources  representing 
the vertex operators for the (``semiclassical'') string states are placed at positions 
$\x_1$ and $\x_2$.
 %
 %
 The role of matching onto   source terms is to relate the parameters  of 
 the semiclassical  solution to the quantum numbers ($\D, J,...$)  that label  
 the vertex operators.\foot{Note that the transformation from a cylinder to 
 the complex plane 
 is not  essential  if we are interested only in the value of a 
  correlator of integrated vertex operators.} 
 Then,  using the massless scalar  operators like \rf{vew} or \rf{kz}
 with $J= \sql \k \gg 1, \ \D=J $, the four-dimensional and two-dimensional 
 conformal invariances imply that, in general, the two-point function 
 should have the form 
 \be 
\bra  V_{J} (\rx_1) V^*_{J} (\rx_2)\ket \sim 
 { 1  \ov   |\rx_1-\rx_2|^{2 \D} } \  \int {d^2 \xi_1 d^2 \xi_2 \ov  |\xi_1 - \xi_2|^4} 
     \ . \la{red}
\ee 
 The semiclassical trajectory \rf{ge} is consistent with the special choice 
 of $\rx=(-1,0,0,0), \ \rx'=(1,0,0,0)$; the divergent 2d ``Mobius'' factor 
 should cancel against  the standard 
 normalization of the string path integral \ci{but}.

To apply this method to the two-point function 
$\bra  V_{S} (\rx_1) V^*_{S} (\rx_2)\ket $ of operators 
 \rf{mml},\rf{jk}
\be
\label{spi}
V_{S} (\rx) &=&
c  \int d^2 \xi \ \big[z(\xi) + z(\xi)^{-1}  (x(\xi)-\rx)^2\big]^{-\Delta} \ 
\big[ \partial \rY_{x}(x(\xi)-\rx) \  \bar \partial \rY_{x}(x(\xi)-\rx)  \big]^{S/2}  
\\
\rY_x(x)&=& Y_1(x) + i Y_2(x) = { x_1 + i x_2 \ov z}  \ ,  
\nonumber
\ee
we  should  consider the limit of $ \D \sim S \sim \sql \gg 1 $, with 
$ \S= { S \ov \sql} $ being large. 
As was  demonstrated in \ci{but} (see also \ci{tsev,b1}) 
the semiclassical trajectory saturating this  two-point correlator is  
equivalent to the conformally transformed \rf{pu} euclidean continuation  
of the asymptotic large spin limit \ci{ft2,ftt} of the  spinning folded string solution 
in $AdS_3$, i.e. 
\be
t=\kappa \tau\ , \ \  \quad \phi\equiv \phi_1=  \kappa \tau\,,\ \  \quad 
\rho=\mu \sigma\ ,
 \ \ \ \ \ \ \ \k=\mu \approx { 1 \ov \pi} \ln \S \gg 1  \ . 
\label{bbb}
\ee
The background  \rf{bbb}  approximates the exact  elliptic function solution  
\ci{gkp2} in the  limit $\k,\mu \gg 1$ on the interval $\s \in[0, {\pi\ov 2}]$; 
to obtain the formal periodic solution on $ 0 < \s \leq 2 \pi$ one  needs to combine
together   four   stretches  $\rho=\mu \sigma$ of the folded string.
  
%
%
 
In the embedding coordinates, the formal euclidean continuation of 
this solution is\foot{Here we depart from the notation in \ci{but} in 
that we do not change the sign of $\phi$ at the same time as 
doing the  euclidean  continuation.}
\bea
&&Y_5=\cosh(\k \t_e)\cosh(\mu \s)\,, \qquad 
Y_{0e}=\sinh(\k \t_e)\cosh(\mu \s)\,,\qquad  Y_4=0 \ ,
\nonumber \\
&&Y_1=\cosh(\k \t_e)\sinh(\mu \s)\,, \qquad 
Y_2= -i \sinh(\k \t_e)\sinh(\mu \s) \ ,  \qquad  Y_3=0 \ .
\label{lll}
\eea
In   Poincar\'e coordinates \rf{yy} this becomes  
 \bea && z=\frac{1}{\cosh(\kappa \tau_e) \cosh(\mu \sigma)}\,,\ \  \ \ \ \ \ \ \ \ 
 \xe  =\tanh(\kappa \tau_e)\,,\la{yji}\\
   &&  x_1= \tanh (\mu \s) \ , \ \ \  \ \ \ \ \ \ 
  x_2 = -i \tanh (\mu \s)\ \tanh(\kappa \tau_e)\,,\la{jji}\\  
&&
  x_\pm\equiv x_1 \pm  i x_2 = r e^{\pm i \phi} 
 =\frac{\tanh (\mu \s)}{\cosh(\k \t_e)} \   e^{ \pm  \k \tau_e}\, , 
\label{vvw}\\ 
&& z^2   + \xe^2  + x_1^2 + x_2^2 =1   \ . \la{vv}
\eea
While in Poincar\'e coordinates in Lorenzian signature the string moves 
towards the center of AdS, rotating and stretching,  after the euclidean 
continuation the resulting  complex  world surface  
described by  \rf{lll} approaches the boundary ($z \to 0$)  at 
$\tau_e \to \pm \infty$  at 
  $\xe (\pm \infty) = \pm 1$
and  ``light-like'' lines in the  (complex)  $(x_1,x_2)$ plane:
 \be 
&& \tau_e \to  + \infty \ : \ \ \ \   z\to 0 \ , \ \ \xe \to 1  \ , \ \ 
 \ \ \  \,  x_+  \to  {2} \tanh (\mu \s) \ ,\ \ \   x_-  \to 0 \ ,   \la{kp}\\
 && \tau_e \to  - \infty \ : \ \ \ \   z\to 0 \ , \ \ \xe \to -1  \ , \ \ \ 
 x_+  \to  0 \ , \ \ \ \ \ \ \ \ \  \ \ \ \ \ \ \,
 x_- \to  {2 } \tanh (\mu \s)     \la{kip}\ee
The   radius  $r= \sqrt{1 -\xe^2 - z^2} $ 
  in the $(x_1,x_2)$ plane    goes to zero  at the boundary 
  while the angle $\phi$ in \rf{vvw} goes to $\pm i \infty$.

 Note that  the fact that this classical solution 
 does not simply end at two points at the boundary does not represent
  a problem. In general, we are supposed to start with 
  two vertex  operators \rf{jk} parametrized by some arbitrary 
  points $\rx_1$ and $\rx_2$ (which are also the points where dual gauge theory
  operators are inserted  in the SYM  correlator corresponding  to \rf{re}) 
  and then find the classical string trajectory ``sourced''
  by such operators. As was shown in \ci{but}, 
  doing this for the choice of $\rx_1=(1,0,0,0)$ and $\rx_2=(-1,0,0,0)$
  (or similar choice  related  by rescaling and translation, see
  footnote \ref{foott})
  leads to the stationary-point solution \rf{yji}--\rf{vv}.
  Thus positions of the boundary values  of the 
  classical string coordinates need not, in general, coincide with 
  the positions of the vertex operators 
  $\rx_1$  and $\rx_2$  (though that does happen for
   simple string solutions which are
  point-like in $AdS_5$, cf. \ci{ja,zarr}).





The discussion above generalizes straightforwardly to the  large spin 
operator carrying also large orbital momentum $J= \sql \J$  in $S^5$, 
\be \la{sj}
V_{S,J}(0)  =  \int d^2 \xi \   (\rY_+)^{-\Delta} \ (\rX_x)^J \ 
\big( \partial \rY_x \  \bar \partial \rY_x  \big)^{S/2}
\ . \ee
The corresponding euclidean 
semiclassical solution \ci{ftt} is given  by a generalization of the 
euclidean  continuation of  \rf{bbb} 
\be
&&t_e=\kappa \tau_e\ , \ \ \  \phi=-i \kappa \tau_e \ , 
\ \ \    \quad \rho=\mu \sigma\ , \ \ \  \ \   \vp =- i \nu \tau_e\ , \label{gen}\\
&&  \k= \sqrt{ \mu^2 + \nu^2 } \ ,\ \ \   \ \ \ \mu \approx { 1 \ov \pi} \ln \S  \gg1
  \ , \ \ \  \nu = \J \ .
\label{gena}
\ee
Its  energy is 
\be\la{io}
 E-S= \sqrt{ J^2 + { \l \ov  \pi^2} \ln^2 \S} = {\sql \ov \pi} \sqrt{ \ell^2 + 1} \  \ln \S \ , \ \ \ \ \  
\  \ell \equiv  {\nu \ov \mu}  \ . \ee
 Written in Poincar\'e coordinates, it is   
the  same as  in \rf{yji}--\rf{vv}   with $\k^2 = \mu^2 + \nu^2$  and 
$\vp = -i \nu \tau_e$.  Note that in  the formal limit of $\mu \to 0$ we recover  
the geodesic solution \rf{ge}. We will use this observation to test some of 
the calculations described in the next section.

\renewcommand{\theequation}{4.\arabic{equation}}
 \setcounter{equation}{0}

  \section{Semiclassical computation of  3-point functions of 
  two ``heavy''  and  one ``light''  states}
 
 Let us now apply the strategy described in the introduction to the 
 computation of the leading semiclassical contribution to the 
 correlators like $ \bra  V_{H_1} (\rx_1) V_{H_2} (\rx_2) V_L(\rx_3) \ket$
 where the  ``heavy''  and   ``light'' vertex operators 
 are  among   the  operators discussed   in section 2. 
 Again, since the quantum numbers  of the ``light'' operators 
 are  much smaller than those of the ``heavy'' ones (assumed  to be order $\sql$)
 the ``light''  source terms in the string  equations determining
  the stationary point trajectory 
 can be ignored,  so that  this trajectory 
   should be the same as for the  2-point correlator 
 $ \bra  V_{H_1} (\rx_1) V_{H_2} (\rx_2)  \ket$.
 Then   to compute the above  3-point   function we 
 just need to evaluate  it on a  classical string solution 
 carrying the same quantum numbers as the two heavy operators
 (assumed to be of the same type up to opposite signs of spins or momenta, i.e. 
 conjugate  to each other). 
 
 Given that the $\rx_i$-dependence  of the  correlators 
  like \rf{re},\rf{ko}
 is determined  by the conformal  invariance, it is sufficient to consider a
  special
  choice of the points, fixing, e.g., the position of the 
  ``light'' operator to be at zero,  $\rx_3=(0,0,0,0)$.
 In this  case  the contribution of the ``light'' vertex operator  will be given
   by (see \rf{jk},\rf{gh})
 \be 
 V_L(0) = \bc_\Delta \int d^2 \xi\  (\rY_+)^{-\D_L}\ U[x(\xi),z(\xi),X(\xi)]  \ , \la{jjpl}
 \ee
 where $\bc_\D$  is  a 
 normalization of the ``light''  operator.
 Furthermore,  for  all  simple classical string solutions 
 associated   with  the ``heavy'' operators we will consider below 
 will have  the following property (cf. \rf{yy},\rf{ge},\rf{lll}): 
 \be \la{hlo}
 z^2 + x_m x^m =1 \ , \ \ \ \ \  {\rm i.e.}  \ \ \ \ \ 
 Y_4=0 \ , \ \ \ Y_5=\rY_+= z^{-1}   \ . \ee
 Then the leading  semiclassical contribution  to the 3-point function 
 will be given simply by (we assume
  $\D_{H_1}=\D_{H_2} \gg  \D_L\equiv \D $) 
 \be 
  \bra  V_{H_1} (\rx_1) V_{H_2} (\rx_2) V_L(0) \ket \sim 
  \int d^2 \xi\  z_{cl}^{\D}\ U[x_{cl}(\xi),z_{cl}(\xi),X_{cl}(\xi)]  \ , \nonumber \ee
 where  the subscript
  $_{cl}$ on the  arguments of $U$ emphasizes 
 that they are given by
 %
 the classical 
 solution saturating the 2-point correlator  of the ``heavy'' operators.\foot{The 
 corresponding worldsheet can be pictured as connecting  the  $\rx_1$  and $\rx_2$
  points
 with  the role of the  third ``small''
operator being to  connect it also to the point $\rx_1=0$ (see \ci{zarr}).}

  As follows from \rf{ko}, the  (normalized) 
  structure coefficients $C_{123}\sim { \C_{123}\ov \C_{12}} $ 
will then be given by 
 \be 
  C_{123} = {\bra  V_{H_1} (\rx_1) V_{H_2} (\rx_2) V_L(0) \ket 
   \ov \bra  V_{H_1} (\rx_1) V_{H_2}(\rx_2)\ket } 
    =\ \Big({|\rx_1| \ |\rx_2|\over |\rx_1 -\rx_2|}\Big)^{\Delta} \ 
     \bc_\D \ 
  \int d^2 \xi\  z_{cl}^{\D}\ U[x_{cl}(\xi),z_{cl}(\xi),X_{cl}(\xi)]  \ . \la{jwpl} \ee
 In such a ratio the (divergent) ``Mobius" factor \rf{red} cancels out as well, 
 guaranteeing a finite result.
 In addition, since dependence on $\rx_1$ and $\rx_2$ should be fixed by conformal
 invariance, 
 we may consider a special choice of them such that 
 $  |\rx_1|=|\rx_2|=1, \  |\rx_1 -\rx_2|=2$
 (this  will be a consistent choice for solutions we shall
  consider below).\foot{For example, the euclidean $AdS_5$ trajectory corresponding 
  to  a  string moving in $S^5$   may be chosen to  approach the boundary points 
  $\rx_1$ and $ \rx_2$  
 satisfying  $|\rx_1|= 1, \  |\rx_2|=1$. Also, for 
\rf{lll}--\rf{vv}   the  boundary points  are   $
x(\tau_e=\infty) = (1, \tanh \mu \s ,-i\tanh \mu \s, 0), $  and $
x(\tau_e=-\infty) = (-1, \tanh \mu \s,i\tanh \mu \s, 0) $
so that 
$|x(\tau_e=+ \infty)|^2 =|x(\tau_e=- \infty)|^2 =1$.} 
Then \rf{jpl} takes the form 
 \be 
  C_{123} =
     c_\D \ 
  \int d^2 \xi\  z_{cl}^{\D}\ U[x_{cl}(\xi),z_{cl}(\xi),X_{cl}(\xi)]  \ ,
  \ \ \ \ \ \ \ \ \  \ \     c_\D \equiv 2^{-\D}\  \bc_\D   \ .  \la{jpl} \ee
 In what follows we shall omit the subscript ``${cl}$" on the 
 coordinates of the classical solution. Also,   since we are interested just in the value of
 the  integral  in \rf{jpl} we may compute it directly on the 2d cylinder, i.e. before doing
 the conformal transformation \rf{pu}, so that 
 $\int d^2 \xi \to \int^{\infty}_{-\infty}
  d \tae \int^{2 \pi}_0  d \s$.

  Let us now  consider some specific examples corresponding to different choices of 
  the  ``heavy''  and  ``light'' 
  operators, i.e. the choices of  the  classical
   solution  and  of  
  $V_L$ \rf{jjpl} or $U$ in \rf{jpl}.

\subsection{$V_H$ corresponding to  folded  string with large spin in $AdS_5$}

Let us start with the case when  the two 
``heavy'' operators are $V_{S,J}$ and $V_{-S,-J}$  in \rf{spi},\rf{sj}
with $S= \sql \S,  \  J= \sql \J$ and $  \ln \S \gg 1,$  so that  the corresponding
semiclassical trajectory is directly  related to the  large spin solution 
\rf{lll}--\rf{gen}.

\subsubsection{$V_L$ as dilaton  operator}

If we choose $V_L$ to be  the dilaton operator \rf{vew} then  
the 3-point correlator  \rf{jpl} takes the form
\be 
 && C_{123} 
    =  c_\D \ \int^{\infty}_{-\infty}
  d \tau_e \int^{2 \pi}_0  d \s\ 
    z^{\D}\   U \ , \la{iii}\\
    &&   U= (\rX_x)^{j} \  
    \Big[ z^{-2} (\del x_m \bd x^m  + \del z \bd z)  +  \del X_k \bd
    X_k \Big]   \ , \ \ \ \ \ \ \D= 4 + j \ .  \la{upl} \ee
Here  we  denoted  the (fixed)  KK momentum of the dilaton by $j$  to distinguish it from 
the (large)   angular momentum $J$ of the ``heavy'' operators.
In this case the  momenta of the two ``heavy'' operators 
should  be, in fact,   $J$ and $-J-j$ to satisfy the momentum 
conservation but as in \ci{zarr}  we shall formally 
  ignore this as $J \gg j$.
   The normalization constant $c_\Delta$ 
  of the dilaton vertex operator  can be  computed 
  following  \cite{bcfm,zarr}:\foot{We followed the procedure of ref. \ci{zarr} 
  using dilaton  field normalization from \cite{bcfm}.}
\be
\bc_\Delta=\bc_{j+4} ={\sql \ov 8\pi N} \  \sqrt{(j+1) (j+2)(j+3)} \ , \ \ \ \ \ \ \ \ 
c_\Delta= 2^{-j-4}\ \bc_{j+4} 
 \ . \la{diel}
\ee
 Let us note  that in the simplest case of the ``heavy'' operators  represented by 
scalar BPS operators corresponding to supergravity modes when  
the classical trajectory is given by \rf{ge}  we  find 
that 
$U= e^{  j\k  \tau_e} \times ( \k^2 - \k^2) =0$ 
so that the 3-point function vanishes identically. 
This agrees with 
the absence of the 3-point couplings  containing 
an odd number of dilatons in the  NS-NS 
sector\foot{We ignore fermions and so do not consider the RR scalar 
operators.}
of the  type IIB supergravity  in the Einstein frame (a similar statement
 is true also in weak-coupling expansion of the dual gauge theory). 

Evaluating $U$ on the large-spin folded string 
classical solution in \rf{gen} 
we get 
\be 
U= e^{   j\nu  \tau_e} \  ( \k^2 \cosh^2 \r  + \mu^2 - \k^2 \sinh^2 \r  -
\nu^2) = 2  \mu^2 \ e^{  j  \nu  \tau_e} \ , \la{hg}   \ee
so that  the integral in \rf{upl}  becomes 
\be 
  C_{123} 
    = 4  c_\D \  \int^{\infty}_{-\infty}
  d \tau_e \int^{{ \pi\ov 2} }_0  d \s\  
   \frac{2\mu^2\ e^{j\nu \tau_e}}
{\big[ \cosh(\mu\sigma)\cosh(\kappa\tau_e)\big]^{\D}}
      \ ,\ \ \ \ \ \ \ \ \ \k^2 = \mu^2 + \nu^2 \ ,  \la{ul} \ee
where we  used that the expression  for $\rho$ in \rf{bbb}
approximates the exact folded solution for $\mu \gg 1$ on the interval
 $(0, { \pi \ov
2})$  and should be combined 4 times to correspond to  a $2\pi$  
periodic  solution. 
While we should eventually take  $\m$ large  we 
shall formally keep it finite at intermediate steps.

Doing the integral over $\s$ and $\tae$ we get\foot{Note that  
the integral over $\tae$ is convergent as $(4+j) \k > j \nu$.} 
\be
&& 
C_{123} =  c_\D \ 2^{j+6
}\, {\mu\ov \k}\  C(j, \mu)   \   B(j,\frac{ \nu}{\kappa})  \ , 
\la{hhh}\\
&& C(j, \mu)=  \sinh({\textstyle{{\pi\ov 2} \mu}}) 
\ {}_2F_1\Big({\textstyle{
\frac{1}{2},\frac{1}{2}(5+j),\frac{3}{2},-\sinh^2({\pi\ov 2} \mu) }}
\Big)   \ , \la{joji} \\
&&
B(j, \frac{ \nu}{\kappa}) = \frac{{}_2F_1(4+j,\  b_+ ,\  b_+ + 1  ,\ -1)}
{ b_+} 
+\frac{{}_2F_1(4+j,\  b_- ,\  b_- +1 ,\ -1)}
{ b_-}   \ , \la{puu} \\
&& b_\pm \equiv  
\frac{1}{2}
\left[4+ j(1 \pm \frac{ \nu}{\kappa})\right]  \ . \no
\ee
Note  that in the formal $\mu\to 0$ limit corresponding to the case 
when the classical trajectory \rf{gen} degenerates into a geodesic 
we recover the vanishing of the 3-point coupling mentioned above
\be
C_{123}\Big|_{_{\mu \to 0}}
=\frac{c_\Delta 2^{j+5} \pi}{(j+2)(j+3)}\;\frac{\mu^2}{\nu}+{\cal O}(\mu^3) \ . 
\ee
Considering the large  spin  or large
  $\mu={1 \ov \pi} \ln \S$ limit  with fixed $
\ell = { \nu \ov \mu} $ (cf. \rf{io})  we may express \rf{hhh} in terms 
of $\S$ and
$\ell$  in the following factorized form 
\be  &&
C_{123} = c_\Delta\; 2^{j+8} \  C(j, \S)\   \td B (j,\ell) \ , \ \ \ \ \ \ 
\td B (j,\ell) = { 1 \ov \sqrt{ \ell^2 +1} }\ 
 B(j,  { \ell \ov \sqrt{ \ell^2 + 1} } ) \ , \la{qqy}
\\
&& C(j, \S)= \ha \S^{1/2} (1- {\S^{-1}})\;
{}_2F_1\Big({\textstyle{
\frac{1}{2},\frac{1}{2}(j+5),\frac{3}{2},-\frac{1}{4} (\S +{\S^{-1}}-2)}}\Big) 
\la{cac} \ . \la{ttt}\ee
%
%
To leading order in the small $\ell$ expansion the ${\tilde B}$ simplifies substantially, 
being expressible only in terms of $\Gamma$ functions. Further expanding at large $\S$
we find that
\be
C_{123}=8c_\Delta \sqrt{\pi}\frac{\Gamma((j+4)/2)}{\Gamma((j+5)/2)}
\left[\sqrt{\pi}\frac{\Gamma((j+4)/2)}{2\Gamma((j+5)/2)}
-\frac{2^{j+4}}{(j+4)\S^{(j+4)/2}}+\dots\right]+{\cal O}(\ell)~~.
\label{dilell0largeS}
\ee

For  $j=0$  or  $\D=4$   we get (recalling  that $\nu=\J$) 
\be
C_{123}= \frac{32 c_\D [2+\cosh(\mu\pi)] \sinh(\mu\pi/2)}{9 \cosh^3(\pi\mu/2) 
\sqrt{1+\ell^2}\,}
= \frac{64 \,c_\D (\S-1)(\S^2+4\S+1)\,\ln \S}{9\pi
     (\S+1)^3  \  \sqrt{\J^2+ {1\ov \pi^2} \ln^2 \S}\,}  \ .  \la{joj}
\ee
In the  large $\S $ limit this becomes 
\be
C_{123} \ \sim \ \frac{
\ln  S}{
\sqrt{ J^2+ {\l \ov \pi^2} \ln^2 S}}  \ . \la{kou}
\ee
Thus  if  $J \gg {\sql \ov \pi} \ln  S$ the 3-point coupling again  vanishes,  
in agreement with the above
 argument that the dilaton does not couple to BPS states.
 If $ J \ll {\sql \ov \pi} \ln  S$  the 3-point
  coupling approaches a constant, which is consistent  with 
  the expectation 
   that the dilaton should generically couple to massive string modes, e.g., 
   via their mass term in a string field theory  action.
   For example,
   adding a massive scalar to a string effective
    action one gets  an exponential
    dilaton
   coupling in the mass term in  the Einstein frame, 
   ${\rm S}=  \int d^{10} x\ \sqrt g  \big( \del^\m \Psi \del_\m\Psi   + M^2 e^{\g \Phi}
   \Psi^2+ ...
  \big)$. Starting  with  such action 
 the 3-point function may be computed using  standard 
  methods \ci{mal,three}, 
  e.g. as in the case when all three modes are supergravity  modes.

The expression in \rf{kou} resembles   the 
$\l$-derivative  of the strong-coupling limit of the 
dimension  of the large spin twist $J$ operator (equal to  the energy of the 
string solution in \rf{io}):
\be 
\l  {\del  \D_{S,J}  \ov \del \l }
=   {\l  \ln^2  S  \ov  2\pi^2 \sqrt{J^2+ {\l \ov \pi^2} \ln^2  S} } + ...  \ ,
\ \ \ \ \ \
\D_{S,J} = S +  \sqrt{J^2+ {\l \ov \pi^2} \ln^2  S} + ... \ .
\la{qww}\ee
Indeed, this relation  should  be  expected in  view of the discussion in section  2.1
(see \rf{zdil}, \rf{diz}). 
Comparing to \rf{kou}, there is, however,  a mismatch in one factor
of $\ln S$. 
This  appears to be depending  on  how  one regularizes 
divergent integrals appearing in the case of insertion of  the dilaton operator
 integrated over 4-space \rf{zdil}. 
 Indeed, to repeat the above computation with  the
 dilaton  vertex operator replaced  by its 0-momentum 
 version, i.e. by the string action, we should omit the $z^\Delta$ factor  in \rf{iii}.
 Then for $j=0$ we get from  \rf{ul} 
 $ 
  C_{123} 
    = 8 \mu^2   c_4 \  \int^{\infty}_{-\infty}
  d \tau_e \int^{{ \pi\ov 2} }_0  d \s\  $.
 Since from the form of the classical solution \rf{lll}--\rf{jji}
 it is clear that the space-time  coordinates depend on $\tau_e$ through 
 $\bar \tau_e= \kappa \tau_e$  it is natural  to introduce a cutoff  $L$  on 
 $\bar \tau_e$. That gives $ C_{123} \sim { \mu^2 \ov \kappa}  L $ 
 which is indeed the same function of $S$ and $J$ as $  \l  {\del  \D_{S,J}  \ov \del \l }
 $ in   \rf{qww}.\foot{
 Note that if we consider unintegrated vertex operators  before dividing over M\"obious group factor 
  then in the  analog of \rf{ko} we will get like in \rf{red} 
   an extra factor of world-sheet distance powers:
$  \bra  \V (\rx_1,\xi_1) \V^* (\rx_2,\xi_2)  \V_3 (\rx_3,\xi_3)  \ket \sim 
  {1 \ov  |\xi_1 - \xi_2|^2 |\xi_1 - \xi_3|^2 |\xi_2 - \xi_3|^2 } 
$
where $\V_3= \V^{(\rm 0-dil)}$  here is the string Lagrangian. 
Integrating this over $\xi_3$  to get insertion of  $V^{(\rm 0-dil)}$ 
or the string action  produces ${1 \ov  |\xi_1 - \xi_2|^4 } \ \ln (a |\xi_1 - \xi_2|) $ 
where $a$ is a world-sheet cutoff. The 
integral of this factor then cancels   against   the normalization to the 2-point function.}

The above relation between the  value of the classical  action on a solution
(with time integral cut off using $t= \kappa \tau$ variable) 
 and 
the derivative of the corresponding classical  energy over string tension for fixed 
spins appears to be quite general (and can be argued for using 
thermodynamical arguments as in \ci{rot}). We shall  see it applying also in the example
discussed in section 4.2.1.

Let us comment also on  the formal
 limit of large  $j$  which  is easy to  analyze by evaluating the
 integral in \rf{ul}  over $\tae$  in a   saddle-point 
approximation  (see  \ci{zarr} fo a similar discussion).\foot{
Unlike the $\tae$ integral, the $\sigma$ integral in \rf{ul} does not posses a real 
saddle point | the integrand is an increasing function, so that we evaluate 
this integral exactly.}
  Rescaling $\tae$  by $\k$ first 
 we end up with (cf. \rf{qqy},\rf{ttt})
\be
&& C_{123} 
=\frac{8 c_\Delta }{\pi\,(1+\ell^2)^{5/2 }}\  {C}(j, \S) \ \ e^{ \,j\,h(\ell)} \ , 
\la{che} \\
&& h(\ell)=-\ha  \Big[ 
\ln (1+\ell^2)+ \frac{\ell}{\sqrt{1+\ell^2}}\ln \frac{\sqrt{1+\ell^2}-
\ell}{\sqrt{1+\ell^2}+\ell}\Big]  \ , \la{chac} 
\ee
where 
\be\la{kpoi}
h (\ell)=
\left\{
\begin{array}{ll}
\ha \ell^2-\frac{5}{12}\ell^4+\dots&, \ \ \ \ \  \ell\rightarrow 0
\cr
\ln 2 -\ha \left(\frac{1}{2}+\ln 2+\ln \ell\right)\frac{1}{\ell^2}
+\dots&, \ \ \ \  \ell\rightarrow\infty
\end{array}
\right.
~~.
\ee
Taking  $\S\rightarrow\infty$ in  ${C}(j, \S)$ given in 
 (\ref{ttt}) and then expanding it at large $j$ 
 we find\foot{Numerical analysis suggests that the  
 result below  holds also at finite $\S$.}
\be
{C}(j, \S)\Big|_{\S\rightarrow\infty} =
%
 \frac{2^{j+2}[\Gamma((j+4)/2)]^2}{\Gamma(j+4)}+{\cal {O}}(\S^{-1})
=\sqrt{\frac{\pi}{2j}}+{\cal O}(j^{-3/2},\S^{-1})~~.
\la{kipu}\ee
Since here  $j$ dependence is not exponential, it is subject
 to  corrections coming from   fluctuations
around the  saddle point of the $\tae$ integral.

\subsubsection{$V_L$ as superconformal  primary scalar operator}

Let us now turn to  the case when the ``light'' operator 
 is   another massless  scalar vertex 
operator in \rf{kz}. In the case when the classical solution is a BMN geodesic or
a folded spinning string in $S^5$ 
representing  a  massive  string mode 
 with spins $(J_1,J_2)$  similar computation was done recently in
\ci{zarr}.
Here we will consider  the case of the 
large spin folded string  solution in $AdS_5$.

In the case of \rf{kz} the factor 
 $U$ in \rf{jpl}  evaluated on the large spin solution \rf{yji},\rf{jji},\rf{gen}
takes the form (we again use $ j$ for the $S^5$
momentum of the ``light''  operator so that here $\D= j$; \   cf. \rf{hg})
\be 
U= e^{   j\nu  \tau_e} \ 
 \Big[ \k^2 ( {2 \ov  \cosh^2 (\k \tae )}  -1 )  + \mu^2
   ( {2 \ov  \cosh^2 (\m \s)   }  -1 )   + \nu^2 \Big] 
   \ . \la{hjg}   \ee
Then   the integral in \rf{upl}  becomes 
\be 
  C_{123} 
    = 4  c_\D \  \int^{\infty}_{-\infty}
  d \tau_e \int^{{ \pi\ov 2} }_0  d \s\  
   \frac{2\ e^{ j\nu \tau_e}}
{\big[ \cosh(\mu\sigma)\cosh(\kappa\tau_e)\big]^{\D}}
      \Big[ {\k^2  \ov  \cosh^2 (\k \tae)  } 
  - \mu^2 \tanh^2 ( \mu \s) \Big]          \ .  \la{ula} \ee
In each term  the $\tae$ and $\s$ integrals  factorize. 
Even in the large $\mu$ limit  
the result is a relatively complicated function of $\ell$ and $j$
which can be analyzed in various limits. 

In the large $\ell={\nu \ov \mu}={\pi \J \ov \ln \S}$  limit we find 
 \be
 C_{123}= c_\Delta 2^{j+2}\sqrt{\pi} 
 \frac{j-1}{j}\,\frac{\Gamma(j/2)}{\Gamma((j+3)/2)}\,\ell+{\cal O}(\ell^{-1}) \ ,  
 \la{q2}
 \ee 
 i.e.  the 3-point function scales proportionally  to $J$.

 The leading term in the small $\ell$ expansion for general $\mu$ is
 \be
 &&C_{123}=8c_\Delta\sqrt{\pi}\frac{\Gamma((j+2)/2)}{j\Gamma((j+3)/2)}
 \Bigg[
 \ \ 
 \frac{1}{\cosh^{(j+1)/2}(\mu\pi/2)} 
 \no \\
 &&\!
 +\ (j-1)\,{}_2F_1\big({\textstyle{\ha , \frac{j+1}{2}, \frac{3}{2},
 -\sinh^{2}(\mu\pi/2)}}\big)
 \Bigg]\sinh(\mu\pi/2)+{\cal O}(\ell)
 \ . \la{jpu}
 \ee
 Taking  $\mu= {1 \ov \pi} \ln \S $  large,  the leading term
 here is 
 \be
C_{123}= 8c_\Delta\sqrt{\pi}\frac{\Gamma((j+2)/2)}{j \Gamma((j+3)/2) }
 \Big[
 \sqrt{\pi}\frac{(j-1)\Gamma(j/2)}{2\Gamma((j+1)/2)}+\frac{ 2^j}{j \ \S^{j/2}}+\dots
 \Big] \ , \la{pqp}
 \ee
 i.e. the 3-coupling approaches a constant. 
 It is interesting to notice that, up to the different
  normalization constant $c_\D$,
 a simple rational dependence on $j$ and the replacement $j\rightarrow j+4$ 
 (which accounts for the different 
 dimension $\Delta$ 
  of the light operator), the $\Gamma$ functions 
 appearing in the leading term are the same as those appearing in 
 eq.~(\ref{dilell0largeS}). 
 The additional rational dependence on $j$ disappears in the large $j$ limit.
The transformation $j\rightarrow j+4$ suggests that the 
3-point function coefficient
is more naturally expressed in terms of the dimension 
of the light operator rather than 
its R-charge $j$; it also points towards  a certain degree of
 universality of $C_{123}$, with only a weak dependence on the light state.

Note that if we formally take the  small $\mu$ limit  for fixed $\ell$ and $j$ 
we get 
 \be
&& C_{123}=\mu\;\frac{2^{j+4}\pi (1+\ell^2)c_\Delta}{4\ell^2(1+j)+(2+j)^2}
 \cr
&&\times  \Big[
(\sqrt{1+\ell^2}(2+j)+\ell j)\,
 {}_2F_1({\textstyle{ 2+j, \frac{1}{2} (2+j-\frac{\ell j}{\sqrt{1+\ell^2}}), 
                     \frac{1}{2}(4+j-\frac{\ell j}{\sqrt{1+\ell^2}}), -1}})
 \cr
&& +\  
(\sqrt{1+\ell^2}(2+j) -\ell j)\,
 {}_2F_1({\textstyle{2+j, \frac{1}{2} (2+j+\frac{\ell j}{\sqrt{1+\ell^2}} ), 
                     \frac{1}{2} (4+j+\frac{\ell j}{\sqrt{1+\ell^2}}), -1}}) 
 \Big]~~.
 \la{nuy}
 \ee
 Taking  then 
 $\ell\rightarrow 0$  we find  a nonvanishing result:
\be
C_{123}=\mu \Big[\frac{8c_\Delta\pi^{3/2}\,
\Gamma((j+4)/2)}{(j+2)\Gamma((j+3)/2)}+{\cal O}(\ell)\Big] \ . \la{ooo}
\ee
 This term arises entirely from the contributions 
 that would  vanish if the limit $\mu\rightarrow 0$  
 were taken directly in the integrand of eq.~\rf{ula}.
The limit $\ell\rightarrow\infty$ of  \rf{nuy} leads to 
\be
C_{123}=  \frac{2^{j+3}\pi c_\Delta}{j+1}\ \mu \ell \ \left[1+{\cal O}(\ell^{-2})\right] \ . 
\ee
%
Using the normalization constant $c_\Delta$ of the chiral primary scalar 
 operator  ($\Delta=j$) in \cite{zarr,bcfm} we get for  the coefficient 
$c_\Delta$  in \rf{jpl},\rf{ula}\foot{Comparing this to dilaton normalization in
\rf{diel}, note  that the two normalization constants 
$(\bc_\Delta)_{dil}$ and $(\bc_\Delta)_{cpo}$ are similar and 
 match in the large $j$ limit.
 }
 \be
\bc_\Delta=\bc_j=\frac{\sql }{8 \pi N}  (j+1)\sqrt{j} , \ \ \ \ \ \ \ \ \ \ \ \ 
c_\Delta=2^{-j}\bc_j=   \frac{\sql }{8 \pi N} \ 2^{-j}\ (j+1)\sqrt{j} ~~.
\ee
%
%
As a result,  in this limit
\be
C_{123} \to \frac{1}{N} J\sqrt{j} \ . \la{kii}
\ee
We thus  formally recover, as  in  a similar  computation in \ci{zarr},  
the result \ci{Lee}  for the  3-point function of the 
 three BMN-type     operators (here with  $j_1=j_2=J, \ j_3=j$).

\subsubsection{$V_L$ as fixed-spin operator  on leading Regge trajectory }

To explore  the structure of the 
  3-point functions  with   the ``light'' state
being  a massive string state  let us now consider the insertion 
of an operator on the leading Regge trajectory, i.e.   $V_s$ in \rf{mml}. 
We change the notation ($S\to s$, $\D\to \D_s$)  assuming now a   
fixed value of spin $s$
and dimension $\D_s = \sqrt{ 2 (s-2)} \fl +...$  which are much smaller than the
semiclassical parameters ($S, \ \D_S \sim \sql$) of the two ``heavy'' operators 
which  are  taken again to   correspond
to the large-spin folded string  solution \rf{lll}--\rf{gen}. 
We shall  ignore the ``mixing'' terms 
indicated by dots in \rf{mml} so  that the result for $C_{123}$ 
 below will be   qualitative. 

In this case  the value of  $U$ in \rf{jpl}  is  (cf. \rf{hg},\rf{hjg})
\be 
U= (\del \rY_x \bd \rY_x)^{s/2} =  e^{  2 s \k \tae} \
\Big[ \mu^2\cosh^2(\mu\sigma)+\kappa^2\sinh^2(\mu\sigma)\Big]^{s/2}
   \ . \la{ag}  
    \ee
For $\nu=0$  (i.e. $\k=\mu$) this becomes 
\be 
U= \mu^{s}   e^{  2 s \mu \tae} \  \big[ \cosh( 2 \mu\sigma) \big]^{s/2}
   \ . \la{agg}  
    \ee
Doing the integral in \rf{jpl} we conclude 
that for  large $\mu$ 
the 3-point coefficient  scales as
\be \la{beh}
C_{123}  \sim  \mu^{s-2}
 \sim 
   (\ln \S)^{s-2}  \ . 
\ee
It is interesting to note that the 2d operator mixing discussed in section 2.3 
does not alter this behavior.
Indeed, it is not hard to see that each derivative in the vertex operator 
brings in a factor of 
$\mu$ while the integration measure cancels two such factors. Since all operators in 
eq.~\rf{cha} have $s$ derivatives, each of them yields an overall $\mu^{s-2}$ factor.

Let us now estimate the large $s$  behaviour of this correlator.
In the large $\mu = {1 \ov \pi} \ln \S$ limit, this can be easily done
 by evaluating the $\sigma$ and $\tae$
integrals in the saddle-point approximation\foot{Here we first rescale 
the 2d coordinates by $\mu$  and then take $\m \to \infty$; we choose 
the real saddle point for the $\sigma$ integral.}
\be
C_{123}&\approx&\mu^{s-2}c_{\Delta_s} 
\int_{-\infty}^{+\infty} \frac{d\tae\  e^{  2 s \tae}}{\cosh^{\Delta_s} \tae}
\int_{0}^{\infty} \frac{d\sigma\  \cosh^{s/2}(2\sigma)}{\cosh^{\Delta_s}\sigma} 
\ =\  \frac{c_{\Delta_s}}{\pi^{s-2}}\  e^{ H (\S,s)}  \ ,  \la{hhk0} \\
   H&=&(s-2) \ln \ln \S +  h_{\tae}(s) + h_{\sigma}(s) \ , \la{hhk1}
\\[2pt]
h_{\tae}&=&\big(\ha\Delta_s-s\big)\ln\big(1-\frac{2s}{\Delta_s}\big)
          +\big(\ha\Delta_s+s\big)\ln\big(1+\frac{2s}{\Delta_s}\big) \ , 
\la{kopu}\\[2pt]
h_\sigma&=&\ha \Delta_s\ln 2  +  \ha \Delta_s 
\ln  \big(1-\frac{s}{\Delta_s}\big)
-  s\ln\big(\frac{\Delta_s}{s} -1 \big)  \ . 
\la{pqqp}\ee
If we further formally assume that $s$  is as large as $\sql$, then 
$\D_s= \sqrt{ 2(s-2)} \fl + ... $ will  also scale as $\sql$,   so that  the function $H$  in the exponent 
will be  proportional to the string tension, as should be 
expected in a semiclassical limit.

\subsubsection{$V_L$ as singlet massive  scalar   operator}

Let us now show that  a similar result to \rf{beh} is found 
if we choose as the ``light'' operator 
the singlet scalar operator \rf{sii} 
representing a string state at level $r-1$. 
The  advantage over the previous case is 
that here   the leading bosonic part of the  operator 
(with dimension $\Delta_r= 2 \sqrt{  (r-1) \sql } + ...$)
is known explicitly.
 We find that the corresponding   factor $U$  in \rf{jpl} 
evaluated on the large spin solution \rf{gen}  
here is (cf. \rf{ag})
\be \la{kkk}
U= (\del X_k\del X_k  \bd X_\ell\bd X_\ell )^{r/2} = \nu^{2r}  \ . \ee
The simplicity of this result  
is a consequence of the special structure of the singlet  operator \rf{sii}
already mentioned in section 2.4: it is built out of chiral components 
of the  $S^5$ sigma model stress tensor 
  which enters the Virasoro conditions. 
This means  that the same constant expression  \rf{kkk}
will be found for any classical  solution describing a string  moving 
non-trivially  in 
$AdS_5$ with its center of mass  orbiting big circle in $S^5$. 
If instead of \rf{sii}  we   consider the $AdS_5$ counterpart
 of this operator  given in  \rf{siii}
  we get the same 
result as in  \rf{kkk} 
\be \la{mmm}
  U=   (\d Y_M \d  Y^M \bd Y_K \bd  Y^K)^{r/2} = \nu^{2 r} \  ,  \ee
since the  Virasoro condition  relates the $AdS_5$ and $S^5$ components of the 
string stress tensor.

Doing the integral in \rf{jpl} we find (cf. \rf{qqy})
\be
&&C_{123}= c_{\D_r} \  C(r,\S) \   \hat B(r,\ell)  \  , \la{pw}
\\
&&C(r,\S)=\ (\ln \S)^{2r-2}\ (\S^{1/2}-\S^{-1/2})\ 
{}_2F_1\Big({\textstyle{ \frac{1}{2},\frac{1}{2}(\Delta_r+1),\frac{3}{2},-
\frac{1}{4}(\S- 2   + \S^{-1}) }} \Big) \ , \la{pt}
\\
&&\hat  B(r,\ell)=
\frac{2^{\Delta_r-2} \ [\Gamma(\Delta_r/2)]^2   }{\pi^{2r-2}\  \Gamma(\Delta_r) 
}
\ \frac{  \ell^{2 r}  }{  \sqrt{1+\ell^2}    } 
\ . \la{qqj}
\ee
In the large spin limit with fixed $\ell$ we get:
\be
C_{123}  \sim  
(\ln \S)^{2r-2}\,\frac{\ell^{2r}}{\sqrt{1+\ell^2}} \ \sim \ \  {  J^{2r} \ov 
\ln S \ \sqrt{ J^2 + { \l \ov \pi^2} \ln^2   S} }  \ ,  \la{kqwk} 
\ee
where we ignored an 
 overall $\D_r$ dependent factor 
 ($\frac{\sqrt{\pi}2^{\Delta_r-2}\Gamma^3(\Delta_r/2)}{\pi^{2r-2}
 \Gamma((\Delta_r+1)/2)\Gamma(\Delta_r)}$)
  that may  be  cancelled  against the 
  normalization $c_{\D_r}$ of the ``light'' vertex operator.

For fixed $\ell$ 
the resulting behaviour of the 3-point function with large spin $S$ is
thus  the same as 
in \rf{beh}: $\ln S$ in power of  the value of the  string level.
For example,  
for the  singlet operator from the  first excited level $r=2$ 
(which should  be dual to a member of Konishi multiplet \ci{rt9}) 
we get  $C_{123}  \sim \ln^2 S\  \frac{\ell^{4}}{\sqrt{1+\ell^2}} $. 
At the same time, this  3-point correlator  vanishes in the $\ell\to 0$ limit (i.e. 
if $ {\sql \ov \pi} \ln S \gg J$), which was not the case for the fixed-spin $s$ 
operator in \rf{agg},\rf{beh} (this vanishing 
follows directly from the special structure 
of the singlet operator in    \rf{kkk}).



\subsection{$V_H$ corresponding to
  ``small'' circular
  string  solution  in $S^5$  with 
$J_1=J_2\not=J_3$ }

Let us now  consider the case of a  
  ``heavy'' state  for which the semiclassical approximation
to the 
two-point function is 
dominated  by the  rigid  circular 
string solution in $S^5$ with three  spins $J_1=J_2$ and  $J_3$  \ci{ft3}.
%
 We shall consider the ``small-string'' branch of this solution 
 which admits the small-spin limit  (see also \ci{rt9})
  \be 
&&  t= \k \tau  , \ \ \   X_1 + i X_2 = a\ e^{i w \tau + i \s}  , \ \ \ 
   X_3 + i X_4 = a\ e^{i w \tau - i \s} , \ \ \ \
   X_5 + i X_6 = \sqrt{ 1 - 2 a^2}\  e^{i \nu  \tau } \no \\
&&  w= \sqrt{1 + \nu^2} , \  \ \ \k= \sqrt{ 4 a^2 + \nu^2}   , \ \ \ 
J_1 =J_2 = J=  \sql a^2 w  , \ \ \ J_3 = \sql  ( 1 - 2 a^2) \nu  \ .  
\la{smat}
\ee
Transforming $ t= \k \tau$ into  Poincar\'e coordinates  and rotating to Euclidean 
signature as in \rf{ge}
\be 
&&z=\frac{1}{\cosh(\kappa \tau_e)}\,, \qquad \xe =\tanh(\kappa \tau_e)\,,\ \ \ \ \ 
 \   \tau_e =  i \tau \  , 
\label{gej}  
 \ee
we  get a (complex)  background  for the $X_k$ coordinates in terms of $\tau_e$ and $\s$
which  should  then be substituted into the integrand  in \rf{jpl}. 
\foot{Let us mention that a similar semiclassical computation 
of the 3-point   string amplitude  involving two  states corresponding to $J_1=J_2\gg 1 $ circular  spinning string  and a graviton as a light operator 
was first  considered in flat space in \ci{russo}.
 There
 it was checked
that the result of the semiclassical calculation agrees with the large spin limit of the exact  correlation 3-point correlation  function.}

\subsubsection{$V_L$ as dilaton  operator}

In this case the integral  in \rf{upl} is found to be (here 
$\D=4+j$  and the integral over $\s$ is trivial as $z$ in \rf{gej} depends only on $\tae$) 
\be 
 C_{123} 
    = 2\pi   c_\D \ \int^{\infty}_{-\infty}
  d \tae\  
   \frac{  (1 - 2 a^2)^{j/2} \  e^{  j\nu\tau_e}  }
{\big[ \cosh(\kappa\tau_e)\big]^{\D}} \times  4 a^2  
    \ .
\la{kq}  \ee 
This expression  vanishes as it should
 in the $a \to 0$ limit when the ``heavy'' state becomes 
 a BMN state. 
The integral over $\tau_e$ is convergent  since 
$ (4 + j) \k  > j \nu$. Explicitly, we get
  \be 
 C_{123} 
    = 
    c_\D  8 \pi a^2 (1 - 2 a^2)^{j/2} \   
     \ \int^{\infty}_{-\infty}
  d \tae 
   \frac{   e^{  j \nu\tae}  }
{\big[ \cosh(\kappa\tae)\big]^{\D}} 
    \ .
\la{kqq}  \ee 
For $\nu=0$,  i.e.for the ``small''  2-spin classical trajectory for which 
$J= \sql \J, \ \J= a^2, \ \D_J= \sql  \k= 2 \sqrt{ J}\fl  $,  we get,  
using that $\Delta=j+4$, 
\be
(C_{123})_{\nu=0}=c_\Delta 8\pi^{3/2} \frac{\Gamma((j+6)/2)}{(j+4)\Gamma((j+5)/2)}
\   a(1-2 a^2)^{j/2} \ \sim \   \sqrt{ J} \ \left(  1 - 2 {J\ov \sql}\right)^{j/2} 
\ . \la{ooss}
\ee
For  $\nu\not=0$ the result is:
\be
C_{123}&=& 2^{j+7} \pi c_\Delta \frac{a^2(1-2a^2)^{j/2}}{\sqrt{\nu^2+4a^2}}
\Big[\frac{ {}_2F_1\big(4+j, \frac{b_-}{2}, 1+\frac{b_-}{2},-1\big)    }{b_-}
       +\frac{ {}_2F_1\big(4+j, \frac{b_+}{2}, 1+\frac{b_+}{2},-1\big)        }{b_+}
\Big] \no 
\\
b_{\pm}&=&4+j\pm \frac{j\nu}{\sqrt{\nu^2+4a^2}} \ . \la{kkkk}
\ee
 Setting here  $j=0$ we get 
\be
(C_{123})_{j=0} = {\tes{\frac{32}{3}}} \pi c_\Delta \ {a^2 \ov  \sqrt{4 a^2+\nu^2}}~ \ . \la{jhd}
\ee
For $a={1\ov \sqrt{2}}$ when the solution \rf{smat} 
reduces to the ``large'' circular solution with  $J_1=J_2, \ J_3=0$ 
and $\D_J= \sqrt{4J^2 + \l}$ \ci{ft3} 
we find\foot{For $j\ne 0$ the limit $a\rightarrow
\frac{1}{\sqrt{2}}$ in \rf{kkkk} yields  vanishing result.}
\be
C_{123}=  {\tes{\frac{16}{3}}}\pi c_\Delta\ \frac{1}{\sqrt{1+w^2}}
  =  {\tes{\frac{16}{3}}}\pi c_\Delta \       \frac{\sql }{\sqrt{4J^2+\lambda}} \ . \la{lpe}
\ee
We observe that like in \rf{kou}, \rf{qww} (and in agreement with the general 
discussion in  section 2.1) 
 this expression  is 
proportional to  the $\l$-derivative of the dimension $\D_J$ of the ``heavy'' state, 
$ \l { \del \ov \del \l}  \D_J =  \frac{\l }{2 \sqrt{4J^2+\lambda}} $. 
The same   result was found in  \ci{cost}  using somewhat different approach.



\subsubsection{$V_L$ as singlet  massive  scalar  operator}

In the  case of the operator in  \rf{sii}  the value of $U$ in \rf{kkk} 
is $ \k^{2r}$ 
 and thus the 
analog of the integral in \rf{kq} is 
\be 
 C_{123} 
    = 2\pi c_\D \ \k^{2 r}   \ \int^{\infty}_{-\infty}
  d \tau_e \  
   \frac{  1 }
{\big[ \cosh(\kappa\tau_e)\big]^{\D_r}}  \ \sim \  \k^{2 r -1 }
    \ .
\la{kqi}  \ee 
Then for the ``small'' string solution with 
$\J_1=\J_2\equiv \J $  and $\J_3 \to 0$  for which  $\k = \sqrt{ 2 \J} $ 
we find  that  
\be \la{find}
 C_{123}\sim   (\sqrt{J})^{2r- 1 }\sim (\D_J)^{2r- 1 }   \ . \ee 
 We conclude 
  again that the 3-point function scales as a  power of the 
level number of the ``light'' string state. 

As discussed
in \ci{rt9}, the  small string or small $\J= { J \ov \sql}$ limit 
of the  solution \rf{smat} may be used to approximate a string state
with fixed quantum number $J$.  
Then, e.g., for $r=2$ representing a state on the 
first excited  string level 
we get $ C_{123} \sim  ( { J \ov \sql})^{3/2} \sim  \l^{-3/4} 
$, i.e. 
 the 3-point function of such  three massive 
 string states is constant for fixed quantum numbers.

\section{Concluding remarks}

The  general correlation functions of local gauge-invariant operators 
in planar ${\cal N}=4$ SYM theory expanded at strong coupling 
are given  by perturbative string-theory correlators  of 
vertex operators of the dual string states.  Standard arguments suggest that, for
states with large quantum numbers, a semiclassical approach should give reliable 
results.  A  semiclassical limit 
of a correlation function should  be 
determined  by   a stationary  point of the
classical worldsheet action with sources corresponding to the relevant vertex
operators.
%
Introducing additional  vertex operators for states with small 
quantum numbers may then  be treated as a perturbation of a lower-point correlation 
function. To leading order, the evaluation of an $(n+m)$-point correlation functions
with $n$ ``heavy" states and $m$ ``light'' states 
amounts to evaluating the product of $m$ 
``light'' vertex operator  factors
 on the classical worldsheet  surface saturating the $n$-point
correlation function of the``heavy" operators.

Using this strategy we analyzed several examples of three-point functions in which  
dimension 
 of the two operators is much larger than that of the third.
  We 
considered the case  when the 
``heavy'' vertex operators correspond to the large folded spinning string in $AdS_5$ 
and also the case  when they correspond to 
 the ``small'' three-spin circular string on $S^5$.
We have found that if the ``light" vertex operator represents  a BPS state,  
the three-point function approaches a   constant as
 the charges of the ``heavy'' states 
are scaled to infinity. We have also discussed certain excited string states as 
``light" states;  
 in particular, we considered 
 states on the leading  Regge trajectory
as well as special singlet states.
In all such cases we found that the three-point function
depends on the semiclassical parameter raised to a power related to the string 
level of the ``light" state.\foot{This behavior is 
  consistent also with that of the correlators involving    BPS
states  which belong to  the string ground state.}


 Let  us now  discuss   possible sources of 
quantum string ($ 1 \ov \sql$) corrections to the three-point function coefficients 
$\CC_{123}$ in \rf{ko}  or $C_{123}$ in \rf{jpl}.  
One source of 
corrections to $\bra V_H(\rx_1)V_H(\rx_2)V_L(\rx_3)\ket$ are  corrections
to the vertex operators  which 
are of two types: (1) corrections to the dimensions of the operators, 
and (2) corrections to the form of the  vertex operators due to  mixing
at higher orders. 
 For the  ``heavy" operators 
 the  former corrections alter only the worldsheet 
 configuration saturating the two-point
function. 
They can be 
 accounted for by simply replacing the semiclassical  parameters 
of the classical solution 
in the expressions derived at  the 
leading order by their quantum-corrected counterparts. 

%
As for the  higher-loop mixing terms  of 2d operators, they  
are  suppressed by a factor of $1 \ov \sql$  without 
additional dependence on the semiclassical parameters.
 Thus, for the purpose 
of finding the leading terms in the
string semiclassical  expansion, such additional
mixings  may be ignored.
Note also  that in the expressions 
in the previous section 
the dependence on the charges of the classical solution is decoupled from 
the dependence on the charges of the ``light" vertex operator, so that 
such corrections will drop out in ratios of correlation functions 
that are independent of the normalizations of vertex
operators.


These quantum  corrections may be given  a simple 2d 
Feynman diagram interpretation. 
Non-trivial contributions   come from contractions involving 
fields from different types of operators. 
The Wick 
contractions between  two ``heavy" operators scale quadratically with some 
large charge while the Wick contractions between one ``heavy" and one 
``light" operator  scale linearly with a large charge. In a semiclassical
 approach the contributions  of the first type are already included in the 
 renormalization of the 
classical solution describing the two-point functions of the ``heavy" operators.
Thus the relevant  one-loop corrections to the three-point function coefficients 
$C_{123}$ 
arise from Wick contractions between one ``heavy" and one ``light" vertex 
operator  and they 
scale linearly with the charges of the ``heavy" vertex operators.


Let us  now comment on the perturbative calculations of 
such  correlation functions in dual  gauge theory. 
Weak coupling calculations of some simple  correlation functions 
 provided some  early tests of the AdS/CFT correspondence. 
While early  calculations focused on
3-point correlators  of BPS operators which may be extrapolated to strong 
coupling, 
recent one-loop calculations  involving non-BPS 
operators \cite{plef} (see also \cite{OkTs, ADGN}) suggest  an interesting relation 
between  the   3-point  coefficients and anomalous dimensions.
Indeed, the one-loop 
correction to the correlation function of two BPS and one Konishi operator \cite{Bianchi} 
is proportional to the anomalous dimension of the Konishi operator.
 If this pattern
persists at higher orders, this three-point function 
 may provide 
an independent determination of the anomalous dimension of the Konishi 
operator  at strong coupling.

Note that  an algebraic Bethe ansatz approach to the diagonalization 
of the spin chain Hamiltonian  provides sufficient information
to evaluate perturbatively the three-point function coefficient, without 
directly resorting to  Feynman diagram approach~\cite{rv} (for a related 
approach using open spin chains see \cite{OkTs}).
For operators dual to ``fast'' strings, described by Landau-Lifshitz type models,
it is possible to do better. In this case  a useful representation for the 
eigenvectors of 
the dilatation operator may be given in terms of coherent states which are,   in turn, 
determined by solutions of the equations of motion of the LL model.\foot{
One-loop corrections to the correlation functions of $SU(2)$-sector operators 
dual to spinning strings, captured by the expectation value of the non-planar 
dilatation operator, have been discussed in \cite{cjjk}.
While extremal correlation functions  where 
the classical dimension of one operator equals the sum of the classical dimensions 
of the other two   are particularly easy to evaluate
(all correlation functions of operators in the $SU(2)$ sector are of this type),  
this approach is not restricted to such correlators.}  
As the Landau-Lifshitz model arises as the fast string limit of the string  
sigma model \ci{kru}, such an approach may provide a relation to the semiclassical methods
used in this paper. For example, a successful extrapolation to strong coupling 
of non-BPS correlation functions may expose non-renormalization theorems
akin to those governing the behavior of certain leading terms in the anomalous dimensions
of ``long'' operators dual to ``fast'' strings.
It may also suggest how to use integrability methods to tackle the problem 
of 3-point function with all three operators being ``heavy''.

As outlined in the Introduction,
a similar  semiclassical approach
may be attempted also for the calculation of higher-point correlation functions.
Unlike three-point functions,  higher-point functions
should have a nontrivial  4d  position dependence.
Some of their  general features like dependence  on large  quantum
numbers  may still  be possible   to analyze.
A  semiclassical  contribution
computed according to the   recipe used here to
the correlation function of two ``heavy" and two ``light" vertex
operators appears
to be given simply by the  value of the product of ``light" vertex
operators on the
world surface saturating the two-point function of the ``heavy"
operators.  It remains  to be seen if it does capture
a dominant  (in large charge, large $\l$) part
of such four-point functions.
\foot{It is interesting to note that if the two  "light" vertex operators are integrated
dilaton vertex operators, the four-point function computed along these lines
reproduces the behavior suggested  (cf. section 2.1)
by the 
 gauge theory analysis,  ${\rm K}_{1234}\propto (\partial_\lambda\Delta)^2$, 
 where $\Delta\gg 1 $ is the dimension
 of the heavy state  (we used  that 
$ (\partial_\lambda\Delta)^2 \gg \partial^2_\lambda\Delta$).
}
Study of such   four-point functions combined with their expected
factorization properties  may also provide information about  other
three-point functions.

\section*{Acknowledgements }

We are very grateful to E. Buchbinder and  K. Zarembo for useful comments on the draft.
AAT also thanks  E. Buchbinder for  discussions of related problems. 
RR's work was supported by the US Department of Energy under contract
DE-FG02-201390ER40577 (OJI), 
the US National Science Foundation under grants PHY-0608114 and 
PHY-0855356 and the A. P. Sloan Foundation.
AAT acknowledges  the hospitality  of the Pennsylvania  State University 
while this  work was in progress. 
%

\bigskip

\end{document}